\documentclass[aps,prd,superscriptaddress,showpacs,preprint]{revtex4}
\usepackage{graphicx, bm}
\usepackage[usenames]{color}
\usepackage{float}
\usepackage{subcaption}

\begin{document}

%\tightenlines
\draft
\title{Sensitivity limits on the weak dipole moments of the top-quark at the Bestest Little Higgs Model }

\author{E. Cruz-Albaro\footnote{elicruzalbaro88@gmail.com}}
\affiliation{\small Facultad de F\'{\i}sica, Universidad Aut\'onoma de Zacatecas\\
            Apartado Postal C-580, 98060 Zacatecas, M\'exico.\\}

\author{ A. Guti\'errez-Rodr\'{\i}guez\footnote{alexgu@fisica.uaz.edu.mx}}
\affiliation{\small Facultad de F\'{\i}sica, Universidad Aut\'onoma de Zacatecas\\
         Apartado Postal C-580, 98060 Zacatecas, M\'exico.\\}

%\textcolor{blue}{}

\date{\today}
%\maketitle

\begin{abstract}
% insert abstract here

We  presented new researches for the sensitivity limits on the weak dipole moments of the top quark-$Z$ boson interactions in the context of the Bestest Little Higgs Model (BLHM). For this purpose, we derive the corresponding Feynman rules. Among the new
contributions, there are those arising from the vertices of scalar bosons, vector bosons and
heavy quarks contribution: $tq_iS_i$, ($S_i=h_0, H_0, A_0, \phi^{0}, \eta^{0}, \sigma, H^{\pm}, \phi^{\pm}, \eta^{\pm}$);
$tq_iV_i$, ($V_i= \gamma, Z, W, Z', W'$); and $Zq_i\bar q_i$  ($q_i=b, t, B, T, T_5, T_6, T^{2/3}, T^{5/3}$).
With these new vertices, we calculate the one-loop contributions to the weak dipole moments $a^W_t$ and $d^W_t$ of the
top-quark in several scenarios. 
We found that with the parameters of the BLHM of $m_{A_0}= 1000$ GeV, $m_{\eta^0}= 100$
GeV, $F=4000$ GeV, $f=[1000, 3000]$ GeV and $\tan\beta=3$, the $a^W_t$ reaches values with a sensitivity of $a^{W}_{t}= 2.49
\times 10^{-4}-1.26 \times 10^{-5}i$, 
$1.26 \times 10^{-4}-5.41\times 10^{-6}i$, where the main contribution comes from the scalars $h_0$ and $H_0$, 
which couples to the top-quark of the Standard Model and  the top-quarks of the BLHM.
These values seem to be out of the reach of the expected experimental sensitivity of present experiments. However, experiments on future colliders are expected to be sensitive enough to measure the AWMDM of the top-quark.
 Our study complements other
studies at the level of one-loop with an extended scalar sector.

\end{abstract}


\vspace{5mm}

\maketitle

\section{Introduction}

The top-quark with $m_t= 172.76\pm 0.30$ GeV~\cite{Data2020} is the most massive of all observed elementary particles in the
Standard Model (SM). The large top-quark mass corresponds to a Yukawa coupling to the Higgs boson close to unity. This suggests
that the top-quark may play a special role within the SM and that its precise characterization may shed light on the electroweak
symmetry breaking mechanism~\cite{PRL81-1998,PRD59-1999}. Because of the top-quark large mass, its couplings are expected to be
more sensitive to new physics Beyond the SM (BSM) with respect to other particles~\cite{Cao:2020npb}. New physics can manifest itself in different
forms. One possibility is that the new physics may lead to the appearance or a huge increase of new types of interactions like
$tH^+b$ or anomalous Flavor Changing Neutral Current $tqg$, $tq\gamma$ and $tqZ$ ($q=u, c$) interactions. Another possibility
is the modification of the SM couplings that involve $t\bar t \gamma$, $t\bar t Z$, $t\bar t g$ and $tW b$ vertices.

The top-quark is a key particle in various extensions BSM and is considered a laboratory for many experimental or simulation aspects
in searches for new physics. In particular, the top-quark anomalous couplings to bosons in the $t\bar t \gamma$ and $t\bar t Z$ vertices,
have made the top-quark one of the most attractive particles for new physics searches. In this regard, the study of the physics of the
$t$-quark by the Tevatron collider at the Fermilab \cite{PLB713-2012, PLB693-2010,PRL102-2009} and the ATLAS and CMS Collaborations \cite{ATLAS-CONF-2012-126,PRL110-2013,EPJC79-2019,JHEP03-2020,Data2020} at the Large Hadron Collider (LHC) has been developed
significantly in recent years and now represents a very active physics program.

One aspect of top-quark physics which is far less explored is interactions with neutral electroweak gauge bosons and the Higgs
boson. The study of anomalous couplings in electroweak corrections is relatively unexplored, and therefore more detailed studies
are  warranted. Sensitivity to these couplings arises in hadronic collisions through the partonic subprocess $q\bar q \to  (Z, \gamma)
\to t\bar t$. At the LHC the $Z$ mediated process is overwhelmed by the strong production mechanism, rendering the tree level
sensitivity vanishingly small. Electroweak loop corrections are another possible source of sensitivity to anomalous couplings.

Currently, little is known about anomalous couplings of the top-quark with the $Z$ boson. There are no direct measurements
of these couplings; indirect measurements, using LEP and SLC data, tightly constrain only the $t\bar tZ$ vector and axial-
vector couplings. On the other hand, the measurement of the $t\bar t Z$ production cross-section at the LHC \cite{ATLAS-CONF-2012-126,PRL110-2013,EPJC79-2019,JHEP03-2020,Data2020} offers a direct test of anomalous
$t\bar t Z$ couplings. The experimental detection of non-zero Anomalous Weak Magnetic Dipole Moment (AWMDM)
or Weak Electric Dipole Moment (WEDM) of heavy fermions such as $\tau$, $b$, and $t$ at the current sensitivity
of the LHC, would be clear evidence of new physics BSM.

With these motivations, we carried out a study on the weak dipole moments of the quark-top in the context of the Bestest Little Higgs
Model (BLHM) \cite{JHEP09-2010}. The purpose of the BLHM is to solve the hierarchy problem without fine-tuning. This is achieved
through the incorporation of one-loop corrections to the Higgs boson mass through heavy top-quarks partners and heavy gauge bosons.
This extension of the SM predicts the existence of new physical scalar bosons neutral and charged $h_0, H_0, A_0, \phi^{0},\eta^{0},
\sigma, H^{\pm}, \phi^{\pm}, \eta^{\pm}$, new heavy gauge bosons $Z', W'$ and new heavy quarks $B, T,T_5, T_6, T^{2/3},T^{5/3}$.
At the one-loop level, the AWMDM $a^W_t$ and WEDM $d^W_t$ of the top-quark are induced via the Feynman diagrams depicted in Fig.~\ref{dipolo},
where $S_i$ represent scalar bosons, $V_i$ neutral and charged gauge bosons and $q_i$ heavy quarks. Therefore, among the new
contributions of the model, there are those arising from the vertices of scalars bosons, vector bosons and heavy quarks
contribution, that is to say, vertices of the form:  $tq_iS_i$, $S_i=h_0, H_0, A_0, \phi^{0}, \eta^{0}, \sigma, H^{\pm}, \phi^{\pm},
\eta^{\pm}$,  $tq_iV_i$, $V_i= \gamma, Z, W, Z', W'$, and $Zq_i\bar q_i$, $q_i=b, t, B, T, T_5, T_6, T^{2/3}, T^{5/3}$,
respectively. With these vertices we calculate the one-loop contributions to the weak dipole moments $a^W_t$ and $d^W_t$ of the top-quark
and in several scenarios with 
$m_{A_0}= 1000, 1500$ GeV, $m_{\eta^0}= 100, 500$ GeV, $F=[3000,6000]$ GeV, $f=[1000, 3000]$
GeV and $\tan\beta=3$.

The paper is structured as follows. In Section II, we give a brief review of the BLHM. In Section III, we present the predictions
of the BLHM on the weak dipole moments of the top-quark $a^W_t$ and $d^W_t$. Finally, we present our conclusions in Section IV.
In Appendix A, we present the complete set of Feynman rules for the study of the weak dipole moments of the top-quark in the
context of the BLHM. 
In Appendix B,  we provide all the numerical contributions of the particles that induce the AWMDM of the top-quark.

\section{Brief review of the Bestest Little Higgs Model}

Various extensions of the SM have been proposed in order to solve the problem of the mass hierarchy. One of the proposed
extensions is the Little Higgs models (LHM)~\cite{Arkani1,Arkani2} that employ a mechanism named collective symmetry breaking. Its main
idea is to represent the SM Higgs boson as a pseudo-Nambu-Goldstone boson of an approximate global symmetry which
is spontaneously broken at a scale in the TeV range.
In these models, the collective symmetry breaking mechanisms is implemented in the norm sector, fermion sector and the Higgs
sector, which predict new particles within the mass range of a few TeV. These new particles play the role of partners of the
top-quark, of the gauge bosons and the Higgs boson, the effect of which is to generate radiative corrections for the mass
of the Higgs boson, and thus cancel the divergent corrections induced by SM particles.
LHM~\cite{Arkani1,Arkani2,Arkani3} on the other hand already have strong constraints from electroweak
precision data. These constraints typically require the new gauge bosons of LHM to be quite heavy \cite{PRD67-2003,
PRD68-2003}. In most LHM, the top partners are heavier than the new gauge bosons, and this can lead to significant fine-tuning
in the Higgs potential \cite{JHEP03-2005}.

An interesting and relatively recent model is the BLHM \cite{JHEP09-2010} overcomes these difficulties by including separate symmetry
breaking scales at which the heavy gauge boson and top partners obtain their masses. This model features a custodial $SU(2)$ symmetry~\cite{Schmaltz:2008vd,Diaz:2001yz},
has heavy gauge boson partner masses above the already excluded mass range, and has relatively light top partners below the
upper bound from fine-tuning.
The BLHM is based on two independent non-linear sigma models. With the first field $\Sigma$, the global symmetry
$SO(6)_A\times SO(6)_B$ is broken to the diagonal group $SO(6)_V$ at the energy scale $f$, while with the second field
$\Delta$, the global symmetry $SU(2)_C \times SU(2)_D$ to the diagonal subgroup $SU(2)$ to the scale $F> f$. In the first
stage are generated 15 pseudo-Nambu-Goldstone bosons that are parameterized as

\begin{equation}\label{Sigma}
\Sigma=e^{i\Pi/f}  e^{2i\Pi_{h}/f}e^{i\Pi/f},
\end{equation}

\noindent where $\Pi$ and $\Pi_h$ are complex and antisymmetric matrices given by

\begin{eqnarray}\label{Pi}
  \Pi=
 \left(
 \begin{array}{c c c}
  i(\phi_a T^{a}_L +\eta_{a} T_{R}^{a})_{4\times 4} & 0 & 0 \\
  0 & 0 &   i\sigma/\sqrt{2} \\
  0 &-i\sigma/\sqrt{2} &  0
\end{array}
 \right),  \hspace{1cm}
 \Pi_h=\frac{i}{\sqrt{2}}
 \left(
 \begin{array}{c c c}
  0_{4\times4} & h_1 & h_2 \\
  -h_{1}^{T} & 0 &   0 \\
  -h_{2}^{T} & 0 &  0
 \end{array}
 \right),
\end{eqnarray}

\noindent where $\phi_{a}$ and $\eta_{a}$ ($ a = 1,2,3 $) are real triplets, $h_{1}$ and $h_{2}$ Higgs vectors as $\bf{4}'$s of $SO(4)$,
and $\sigma$ a real singlet. For Higgs fields, their explicit representation is $h_{i}^{T}=(h_{i1}, h_{i2}, h_{i3}, h_{i4})$,
while $T^{a}_{L, R}$ denote the generators of the group $SO(6)$ which are provided in~\cite{JHEP09-2010}.

Regarding the second stage of spontaneous symmetry-breaking, the pseudo-Nambu-Goldstone bosons of the field $\Delta$ are parameterized
as follows

\begin{equation}\label{Delta}
\Delta=F e^{2i \Pi_d/F},\, \,\, \, \, \Pi_d=\chi_a \frac{\tau^{a}}{2} \ \ (a=1,2,3),
\end{equation}

\noindent where $\chi_a$ represents the Nambu-Goldstone fields and the $\tau_a$ correspond to the Pauli matrices, which are the
generators of the SU(2) group.

\subsection{The scalar sector}

The BLHM Higgs fields, $h_1$ and $h_2$, form the Higgs potential that undergoes spontaneous symmetry-breaking~\cite{JHEP09-2010,Kalyniak,Erikson}:

\begin{equation}\label{Vhiggs}
V_{Higgs}=\frac{1}{2}m_{1}^{2}h^{T}_{1}h_1 + \frac{1}{2}m_{2}^{2}h^{T}_{2}h_2 -B_\mu h^{T}_{1} h_2 + \frac{\lambda_{0}}{2} (h^{T}_{1}h_2)^{2}.
\end{equation}

\noindent The potential reaches a minimum when $m_1, m_2 >0$, while the spontaneous electroweak symmetry-breaking requires that
$B_\mu > m_1 m_2$. The symmetry-breaking mechanism is implemented in the BLHM when the Higgs doublets acquire
their vacuum expectation values (VEVs), $\langle h_1\rangle ^{T}=(v_1,0,0,0)$ and $ \langle h_2 \rangle ^{T}=(v_2,0,0,0)$. By demanding that
these VEVs minimize the Higgs potential of Eq. (4), the following relations are obtained

\begin{eqnarray}\label{v12}
&&v^{2}_1=\frac{1}{\lambda_0}\frac{m_2}{m_1}(B_\mu-m_1 m_2),\\
&&v^{2}_2=\frac{1}{\lambda_0}\frac{m_1}{m_2}(B_\mu-m_1 m_2).
\end{eqnarray}

\noindent These parameters can be expressed as follows

\begin{equation}\label{vvacio}
v^{2}\equiv v^{2}_1 +v^{2}_2= \frac{1}{\lambda_0}\left( \frac{m^{2}_1 + m^{2}_2}{m_1 m_2} \right) \left(B_\mu - m_1 m_2\right)\simeq \left(246\ \ \text{GeV}\right)^{2},
\end{equation}

\begin{equation}\label{beta}
\text{tan}\, \beta=\frac{v_1}{v_2}=\frac{m_2}{m_1}.
\end{equation}

\noindent From the diagonalization of the mass matrix for the scalar sector,
three non-physical fields $G_0$ and $G^{\pm}$, two physical scalar fields $H^{\pm}$
and three neutral physical scalar fields $h_0$, $H_0$ and $A_0$ are generated~\cite{Kalyniak,PhenomenologyBLH}. 
The lightest state, $h_0$, is identified as the
scalar boson of the SM. The masses of these fields are given as

\begin{eqnarray}\label{masaAGH}
m_{G_0}&=&m_{G^{\pm}}=0,\\
m^{2}_{A_{0}}&=&m^{2}_{H^{\pm}} =m^{2}_1+m^{2}_2,\label{mHmas} \\
m^{2}_{h_0,H_{0}} &=& \frac{B_\mu}{\text{sin}\, 2\beta}\mp \sqrt{\frac{B^{2}_{\mu}}{\text{sin}^{2}\, 2\beta} -2\lambda_0 B_\mu v^{2} \text{sin}\, 2\beta +\lambda^{2}_{0} v^{4} \text{sin}^{2}\, 2\beta  } \label{mh0H0}.
\end{eqnarray}

\noindent The four parameters present in the Higgs potential $ m_1,  m_2, B_\mu$ and $\lambda_0 $, can be replaced by another more phenomenologically accessible set. That is, the masses of the states $h_0$ and $A_0$, the angle $\beta$ and the VEV $v$~\cite{Kalyniak}:

\begin{eqnarray}\label{parametros}
B_\mu &=&\frac{1}{2}(\lambda_0  v^{2} + m^{2}_{A_{0}}  )\, \text{sin}\, 2\beta,\\
\lambda_0 &=& \frac{m^{2}_{h_{0}}}{v^{2}}\Big(\frac{  m^{2}_{h_{0}}- m^{2}_{A_{0}} }{m^{2}_{h_{0}}-m^{2}_{A_{0}} \text{sin}^{2}\, 2\beta }\Big),\\
\text{tan}\, \alpha &=& \frac{ B_\mu \text{cot}\, 2\beta+ \sqrt{(B^{2}_\mu/\text{sin}^{2}\, 2\beta)-2\lambda_0 B_\mu v^{2} \text{sin}\, 2\beta+ \lambda^{2}_{0} v^{4}\text{sin}^{2}\, 2\beta  }  }{B_\mu -\lambda_0 v^{2} \text{sin}\, 2\beta},\label{alpha}   \\
m^{2}_{H_{0}} &=& \frac{B_\mu}{\text{sin}\, 2\beta}+ \sqrt{\frac{B^{2}_{\mu}}{\text{sin}^{2}\, 2\beta} -2\lambda_0 B_\mu v^{2} \text{sin}\, 2\beta +\lambda^{2}_{0} v^{4} \text{sin}^{2}\, 2\beta  }, \label{mH0}\\
m^{2}_{\sigma}&=&(\lambda_{56} + \lambda_{65})f^{2}=2\lambda_0 f^{2} \text{K}_\sigma. \label{masaescalar}
\end{eqnarray}

From Eq.~(\ref{masaescalar}), the variables $\lambda_{56}$ and $\lambda_{65}$ represent the coefficients of the quartic potential defined
in \cite{JHEP09-2010}, both variables take values different from zero to achieve the collective breaking of the symmetry
and generate a quartic coupling of the Higgs boson \cite{Kalyniak}.
The BLHM also contains scalar triplet fields that get a contribution to their mass from the explicit symmetry breaking terms in model, as define in Ref.~\cite{JHEP09-2010}, that depends on the parameter $m_4$.

\begin{eqnarray}
m^{2}_{\phi^{0}}&=& \frac{16}{3}F^{2} \frac{3 g^{2}_{A} g^{2}_{B}}{32 \pi^{2}} \log \left( \frac{\Lambda^{2}}{m^{2}_{W'}}\right) + m^{2}_{4} \frac{f^{4}+ F^{4}}{F^{2}(f^{2}+F^{2})},\\
m^{2}_{\phi^{\pm}}&=& \frac{16}{3}F^{2} \frac{3 g^{2}_{A} g^{2}_{B}}{32 \pi^{2}} \log \left( \frac{\Lambda^{2}}{m^{2}_{W'}}\right) + m^{2}_{4} \frac{f^{4}+f^{2}F^{2}+F^{4}}{F^{2}(f^{2}+F^{2})},\\
m^{2}_{\eta^{\pm}}&=&  m^{2}_{4}+ \frac{3 f^{2} g^{2}_{Y}}{64 \pi^{2}}\frac{\Lambda^{2}}{F^{2}},\\
m^{2}_{\eta^{0}}&=&m^{2}_{4}. 
\end{eqnarray}

\subsection{The gauge sector}

In the BLHM the new gauge bosons develop masses proportional to $\sqrt{f^2+F^2}\sim F$. This makes the masses of the bosons
large relative to other particles that have masses proportional to $f$. The kinetic terms of the gauge fields in the BLHM
are given as follows:

\begin{equation}\label{Lcinetico}
\mathcal{L}=\frac{f^{2}}{8} \text{Tr}(D_{\mu} \Sigma^{\dagger} D^{\mu} \Sigma) + \frac{F^{2}}{4} \text{Tr}(D_\mu \Delta^{\dagger} D^{\mu} \Delta),
\end{equation}

\noindent where

\begin{eqnarray}\label{derivadasC}
D_{\mu}\Sigma&=&\partial_{\mu} \Sigma +i g_A A^{a}_{1\mu} T^{a}_L \Sigma- i g_B \Sigma A^{a}_{2\mu} T^{a}_L+ i g_{Y} B^{3}_{\mu}(T^{3}_{R}\Sigma-\Sigma T^{3}_{R}),\\
D_{\mu}\Delta&=&\partial_{\mu} \Delta +i g_A A^{a}_{1\mu} \frac{\tau^{a}}{2}  \Delta- i g_B \Delta A^{a}_{2\mu} \frac{\tau^{a}}{2}.
\end{eqnarray}

\noindent $T^{a}_{L}$ are the generators of the group $SO(6)_A$ corresponding to the subgroup $SU(2)_{LA}$, while $T^3_R$ represents
the third component of the $SO(6)_B$ generators corresponding to the $SU(2)_{LB} $ subgroup, these matrices are provided in~\cite{JHEP09-2010}.
$g_A$ and $A^{a}_{1\mu}$ denote the gauge coupling and field associated with the gauge bosons of $SU(2)_{LA}$. $g_B$ and $A^{a}_{2\mu}$
represent the gauge coupling and the field associated with $SU(2)_{LB}$, while $g_Y$ and $B^{3}_{\mu}$ denote the hypercharge and the field.

When $\Sigma$ and $\Delta$ get their VEVs, the gauge fields $A^{a}_{1\mu}$ and $A^{a}_{2\mu}$ are mixed to form a massless triplet
$A^{a}_{0\mu}$ and a massive triplet $A^{a}_{H\mu}$,

\begin{equation}\label{AA}
A^{a}_{0\mu}=\text{cos}\, \theta_g A^{a}_{1\mu} + \text{sin}\, \theta_g A^{a}_{2\mu}, \hspace{5mm} A^{a}_{H\mu}= \text{sin}\, \theta_g A^{a}_{1\mu}- \text{cos}\, \theta_g A^{a}_{2\mu},
\end{equation}

\noindent with the mixing angles

\begin{equation}\label{gagb}
s_g\equiv \sin \theta_g=\frac{g_A}{\sqrt{g_{A}^{2}+g_{B}^{2}} },\ \ c_g \equiv \cos \theta_g=\frac{g_B}{\sqrt{g_{A}^{2}+g_{B}^{2}} },
\end{equation}

\noindent which are related to the electroweak gauge coupling $g$ through

\begin{equation}\label{g}
\frac{1}{g^{2}}=\frac{1}{g^{2}_A}+\frac{1}{g^{2}_B}.
\end{equation}

After the breaking of the electroweak symmetry, when the Higgs doublets, $h_1$ and $h_2$ acquire their VEVs, the masses
of the gauge bosons of the BLHM are generated. In terms of the model parameters, the masses are given by

\begin{eqnarray}\label{masaBoson}
m^{2}_{\gamma} &=0&, \\
m^{2}_{Z}&=&\frac{1}{4}\left(g^{2}+g^{2}_Y \right)v^{2} \left(1-\frac{v^{2}}{12 f^2} \left(2+\frac{3f^2}{f^2+F^2} \left( s^{2}_g -c^{2}_g \right)^{2} \right)  \right), \\
m^{2}_{W}&=& \frac{1}{4} g^{2} v^{2} \left(1-  \frac{v^{2}}{12 f^2} \left(2+  \frac{3f^2}{f^2+F^2} \left(s^{2}_g -c^{2}_g  \right)^{2}\right)  \right),\\
m^{2}_{Z'}&=&m^{2}_{W'} +  \frac{g^2 s^{2}_W v^4}{16 c^{2}_W (f^2+F^2)} \left(s^{2}_g -c^{2}_g \right)^{2}, \label{mzprima} \\
m^{2}_{W'}&=& \frac{g^2}{4 c^{2}_{g} s^{2}_{g}} \left(f^2+F^2 \right)  - m^{2}_{W}. \label{mwprima}
\end{eqnarray}

The weak mixing angle is defined as

\begin{eqnarray}\label{angulodebil}
s_W&&\equiv\sin \theta_W = \frac{g_Y}{\sqrt{g^2+ g^{2}_Y }}, \\
c_W&&\equiv\cos \theta_W= \frac{g}{\sqrt{g^2+ g^{2}_Y }},\\
x_W&=&\frac{1}{2 c_W} s_g c_g (s^{2}_g -c^{2}_g).
\end{eqnarray}

\subsection{The fermion sector} \label{subsecfermion}

To construct the Yukawa interactions in the BLHM, the fermions must be transformed under the group $SO(6)_A$ or $SO(6)_B$.
In this model, the fermion sector is divided into two parts. First, the sector of massive fermions is represented by Eq.~(\ref{Ltop}).
This sector includes the top and bottom quarks of the SM and a series of new heavy quarks arranged in four multiplets, $Q$, and $Q'$
which transform under $SO(6)_A$, while $U^c$ and $U^{'c}_5$  are transformed under the group $SO(6)_B$. Second, the sector of
light fermions contained in  Eq.~(\ref{Lligeros}), in this expression all the interactions of the remaining fermions of the SM with the
exotic particles of the BLHM are generated.

For massive fermions, the Lagrangian that describes them is given by \cite{JHEP09-2010}

\begin{equation}\label{Ltop}
\mathcal{L}_t=y_1 f Q^{T} S \Sigma S U^{c} + y_2 f Q'^{T} \Sigma U^{c} +y_3 f Q^{T} \Sigma U'^{c}_{5} +y_b f q_{3}^{T}(-2 i T^{2}_{R} \Sigma) U^{c}_{b}+ H.c.,
\end{equation}

\noindent where $ S = \text{diag} (1,1,1,1, -1, -1) $. The multiplets are arranged as follows

\begin{eqnarray}\label{camposf}
Q^{T}&=&\frac{1}{\sqrt{2}}\left( \left(-Q_{a_1} -Q_{b_2}\right), i\left(Q_{a_1} -Q_{b_2} \right),  \left(Q_{a_2} -Q_{b_1}\right), i\left(Q_{a_2} -Q_{b_1}\right), Q_{5},Q_{6} \right),\\
Q'^{T}&=&\frac{1}{\sqrt{2}} (-Q'_{a_1}, iQ'_{a_1},Q'_{a_2},iQ'_{a_2},0,0 ),\\
q_{3}^{T}&=& \frac{1}{\sqrt{2}} (-\bar{t}_L, i\bar{t}_L,\bar{b}_L,i\bar{b}_L,0,0 ),\\
U^{cT}&=& \frac{1}{\sqrt{2}} \left( (-U^{c}_{b_1} -U^{c}_{a_2}), i (U^{c}_{b_1} -U^{c}_{a_2}),  (U^{c}_{b_2} -U^{c}_{a_1}), i (U^{c}_{b_2} -U^{c}_{a_1}), U^{c}_{5},U^{c}_{6} \right),\\
U'^{cT}&=&(0,0,0,0,U'^{c}_5,0),\\
U_{b}^{cT}&=&(0,0,0,0,b^{c},0).
\end{eqnarray}

\noindent The explicit forms of the components of the multiplets are defined in Ref.~\cite{JHEP09-2010}. For simplicity, the Yukawa couplings are assumed to be real $y_1, y_2, y_3$ $\in R$. The Yukawa coupling of the top-quark is defined as

\begin{equation}\label{yt}
 y_t= \frac{3 y_1 y_2 y_3}{ \sqrt{ (y^{2}_1 +y^{2}_2)(y^{2}_1 +y^{2}_3)}}=\frac{m_{t}}{v \sin \beta}.
 \end{equation}

\noindent For light fermions the corresponding Lagrangian is

\begin{equation}\label{Lligeros}
\mathcal{L}_{light}= \sum_{i=1,2} y_u f q^{T}_i \Sigma u^{c}_{i} + \sum_{i=1,2} y_{d} f q^{T}_{i}(-2i T^{2}_{R} \Sigma) d^{c}_i
+\sum_{i=1,2,3} y_e f l^{T}_i (-2i T^{2}_{R} \Sigma) e^{c}_i + h.c.,
\end{equation}

\noindent with

\begin{eqnarray}\label{qligeros}
 q^{T}_i &=&\frac{1}{\sqrt{2}} (-\bar{u}_{iL}, i \bar{u}_{iL}, \bar{d}_{iL}, i \bar{d}_{iL},0,0),\\
 l^{T}_i &=&\frac{1}{\sqrt{2}} (-\bar{\nu}_{iL}, i \bar{\nu}_{iL}, \bar{e}_{iL}, i \bar{e}_{iL},0,0),\\
 u^{cT}_i &=&(0,0,0,0,u^{c}_i,0),\\
 d^{cT}_i &=&(0,0,0,0,d^{c}_i,0),\\
 e^{cT}_i &=&(0,0,0,0,e^{c}_i,0).
\end{eqnarray}

After breaking the electroweak symmetry, the resulting mass terms are expanded by a power series up to $\frac{v^2}{f^2}$
and the mass matrices are diagonalized using perturbation theory.
 The fermion mass eigenstates are calculated under the assumption that $y_{2} \neq y_{3}$, otherwise the masses of $T$ and $T_{5}$ are degenerate at lowest order~\cite{PhenomenologyBLH}:

\begin{eqnarray}
  m^{2}_t &=&y^{2}_t v^{2}_1,\label{mt} \\
  m^{2}_T &=& (y^{2}_1 + y^{2}_2)f^2 + \frac{9 v^{2}_1 y^{2}_1  y^{2}_2  y^{2}_3 }{(y^{2}_1 + y^{2}_2) (y^{2}_2 - y^{2}_3)}, \label{mT} \\
  m^{2}_{T_5} &=& (y^{2}_1 + y^{2}_3)f^2 - \frac{9 v^{2}_1 y^{2}_1  y^{2}_2  y^{2}_3 }{(y^{2}_1 + y^{2}_3) (y^{2}_2 - y^{2}_3)},\label{MT5} \\
  m^{2}_{T_6} &=&m^{2}_{T^{2/3}_b}=m^{2}_{T^{5/3}_b} =y^{2}_1 f^2,\label{mT6} \\
  m^{2}_B & =&(y^{2}_1 + y^{2}_2)f^2,\label{mB}
\end{eqnarray}

\noindent with $v_1=v\sin \beta$ y $v_2=v\cos \beta$.

\subsection{The currents sector}

In this sector the interactions of the fermions with the gauge bosons are determined. The vertices are obtained from the
following Lagrangian \cite{PhenomenologyBLH},

\begin{eqnarray}\label{LbaseW}
 \mathcal{L} &=& \bar{Q} \bar{\tau}^{\mu} D_{\mu}Q + \bar{Q}' \bar{\tau}^{\mu} D_{\mu}Q'- U^{c\dagger} \tau^{\mu} D_{\mu}U^{c}-  U'^{c\dagger} \tau^{\mu} D_{\mu}U'^{c} -  U_{b}^{c\dagger} \tau^{\mu} D_{\mu}U_{b}^{c} +\sum_{i=1,2}  q^{\dagger}_i \tau^{\mu} D_{\mu} q_i   \nonumber \\
 &+& \sum_{i=1,2,3}  l^{\dagger}_i \tau^{\mu} D_{\mu} l_i
 - \sum_{i=1,2,3}  e_i^{c\dagger} \tau^{\mu} D_{\mu} e^{c}_i - \sum_{i=1,2}  u_{i}^{c\dagger} \tau^{\mu} D_{\mu} u^{c}_{i} - \sum_{i=1,2}  d_{i}^{c\dagger} \tau^{\mu} D_{\mu} d^{c}_i,
\end{eqnarray}

\noindent where $\tau^{\mu}$ and $\bar{\tau}^{\mu}$ are defined according to~\cite{Spremier}. The respective covariant derivatives are

\begin{eqnarray}\label{dcovariantes}
  D_{\mu}Q & =&\partial_{\mu}Q+ \sum_{a} (i g_A A^{a}_{1\mu} T^{a}_{L} Q )+ ig_Y B_{3\mu} (T^{3}_{R} +T^{+}_{X} )Q, \\
  D_{\mu}Q'  & =&\partial_{\mu}Q'+\sum_{a} (i g_A A^{a}_{1\mu} T^{a}_{L} Q' )+ ig_Y B_{3\mu} \left(\frac{1}{6} \right)Q',\\
   D_{\mu}U^{c} & =&\partial_{\mu}U^{c}+ \sum_{a} (i g_B A^{a}_{2\mu} T^{a}_{L} U^{c} )+ ig_Y B_{3\mu} (T^{3}_{R} +T^{-}_{X} )U^{c},\\
 D_{\mu}U'^{c} & =&\partial_{\mu}U'^{c}+  ig_Y B_{3\mu} T^{-}_{X} U'^{c},  \\
 D_{\mu}U_{b}^{c} & =&\partial_{\mu}U_{b}^{c}+  ig_Y B_{3\mu} \left(\frac{1}{3} \right) U_{b}^{c},  \\
     D_{\mu}q_i & =&\partial_{\mu}q_i+ \sum_{a} (i g_A A^{a}_{1\mu} T^{a}_{L} q_i )+ ig_Y B_{3\mu} (T^{3}_{R} +T^{+}_{X} )q_i,\\
  D_{\mu}l_i & =&\partial_{\mu}l_i +\sum_{a} (i g_B A^{a}_{2\mu} T^{a}_{L} l_i )+ ig_Y B_{3\mu} T^{3}_{R} l_i,\\
   D_{\mu}e^{c}_i & =&\partial_{\mu}e_{i}^{c} + ig_Y B_{3\mu} T^{e}_{X} e^{c}_i,\\
 D_{\mu}u^{c}_i & =&\partial_{\mu}u_{i}^{c}+  ig_Y B_{3\mu} T^{-}_{X} u^{c}_i,\\
    D_{\mu}d^{c}_i & =& \partial_{\mu}d_{i}^{c}+ ig_Y B_{3\mu} T^{d}_{X} d^{c}_i.\label{dcovariantes70}
\end{eqnarray}

The Feynman rules of the BLHM involved in our calculation are obtained by transforming the gauge eigenstate
in terms of the mass eigenstates for the fermions, the gauge bosons, and the scalars bosons (see Appendix C of Ref.~\cite{Martin:2012kqb}).

\section{Sensitivity limits on the AWMDM $a^W_t$ at the BLHM}

The weak properties of the top-quark appear in the quantum field theory of its interaction with the $Z$ boson. In this regard,
the most general Lorentz-invariant vertex function describing the interaction of a $Z$ boson with two top-quarks can be
written in terms of ten form factors \cite{NPB551-1999,NPB812-2009}, which are functions of the kinematic invariants.
In the low energy limit, these correspond to couplings that multiply dimension-four or-five operators in an effective
Lagrangian, and may be complex. If the $Z$ boson couples to effectively massless fermions, the number of independent form
factors is reduced to eight. In addition, if both top-quarks are on-shell, the number is further reduced to four. In this
case, the $t\bar tZ$ vertex can be written in the form

\begin{eqnarray}\label{verticeZtt}
ie\bar{u}(p') \Gamma^{\mu}_{t\bar tZ}(q^{2}) u(p) &=&ie \bar{u}(p')\big\{ \gamma^{\mu}\left[ F_{V}(q^{2})-F_{A}(q^{2})\gamma^{5}\right] \nonumber \\
&+& i \sigma^{\mu \nu} q_{\nu} \left[ F_{M}(q^{2})- iF_{E}(q^{2})\gamma^{5}\right] \big\}u(p),
\end{eqnarray}

\noindent where $e$ is the proton charge. The terms $F_{V}(0)$ and $F_{A}(0)$ in the low energy
limit are the $t\bar tZ$ vector and axial-vector form factors in the SM, while $F_M(q^2)$ and $F_E(q^2)$ are associated with the
form factors of the weak dipole moments. The latter appear due to quantum corrections and are a valuable tool to study the effects
of new physics indirectly way, through virtual corrections of new particles predicted by extensions of the SM. Another
characteristic of form factors is that they only depend on an independent dynamic variable, $q^2$, where $q = p-p'$
and denotes the incoming moment of the $Z$ boson. 
In this work, the $Z$ boson is off-shell mass since to produce a top-quark pair the $Z$ must necessarily be off the resonance. For this case, the well-known gauge dependence problem arises and occurs when one studies the radiative corrections to fermion-pair production at colliders with center-of-mass energy beyond the mass of the $Z$ boson, that is, $\sqrt{q^{2}}>2m_t$~\cite{Bernabeu:1995gs}. In this case, the pinch technique~\cite{Papavassiliou:1993qe,Cornwall:1989gv,Cornwall:1981zr,Alkofer:2000wg,Papavassiliou:1996zn} can be used to remove the gauge dependence. Our results are useful to determine the effects of new physics that could be potentially very important.
With respect to form factors, $F_M (q^2)$ and $F_E(q^2)$, these are
related to the AWMDM $a^{W}_t$ and the WEDM $d^{W}_t $:

\begin{eqnarray}\label{MDD}
 F_{M}(q^{2})&=&-\frac{a^{W}_t}{2 m_t}, \\
 F_{E}(q^{2})&=&-\frac{d^{W}_t}{ e}.
\end{eqnarray}

It is worth mentioning that, the weak magnetic dipole form factor $F_M (q^2)$ receives contributions at the one-loop level in the SM.
However, there is no such contribution to the weak electric dipole form factor, $F_E(q^2)$ \cite{NPB551-1999}. For this reason, we only
estimate limits on the AWMDM of the top-quark.

\subsection{ Contribution of new scalar bosons, gauge bosons and heavy quarks to the AWMDM of the top-quark}

The AWMDM of top-quark carries important information about interactions with other particles. Their small
magnitude in the SM makes these couplings ideal for probing new physics interactions and for exploring the potential role
of top-quark in electroweak symmetry breaking, which has yet to be elucidated. In this way, the top-quark is expected to
be a window to any new physics at the TeV energy scale. In this subsection, we evaluated the AWMDM of the top-quark in the
context of the BLHM. All the possible one-loop contributions to the $F_M (q^2)$ and $F_E(q^2)$ form factors can be classified
in terms of the two classes of triangle diagrams depicted in Fig.~\ref{dipolo}. From this figure we can see
that $S_{i}$, $V_{i}$ and $ q_{i}$ are  the particles circulating in the loop: $S_{i}$ represents the scalars $h_0$
(SM Higgs boson), $H_0, A_0, \phi^{0}, \eta^{0}, \sigma, H^{\pm}, \phi^{\pm}, \eta^{\pm}$; $V_{i}$ stands for the gauge
bosons $\gamma, Z, W, Z', W'$; and $q_i$ denotes the quarks $b, t, B, T, T_5, T_6, T^{2/3}, T^{5/3}$.
To obtain the amplitude of each contribution we need to know the Feynman rules involved in the diagrams shown in
Fig.~\ref{dipolo}, these vertices are provided in Appendix A.

In the unitary gauge, there are 52 diagrams that contribute to  vertex $t\bar{t}Z$.
We classify these contributions into two categories, which are written in the following compact form:

\begin{figure}[H]
\center
\subfloat[]{\includegraphics[width=5.5cm]{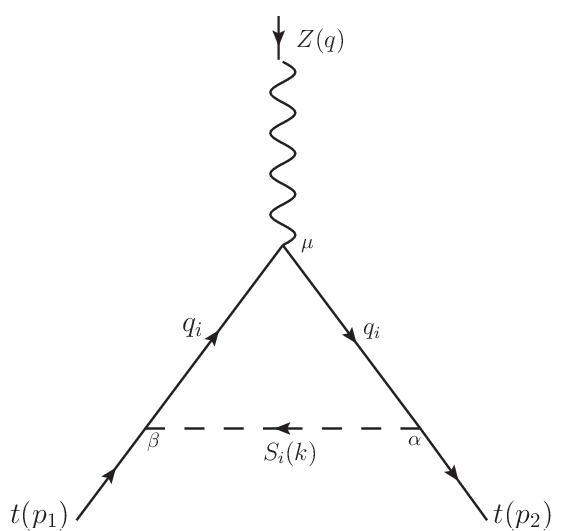}}
\subfloat[]{\includegraphics[width=5.5cm]{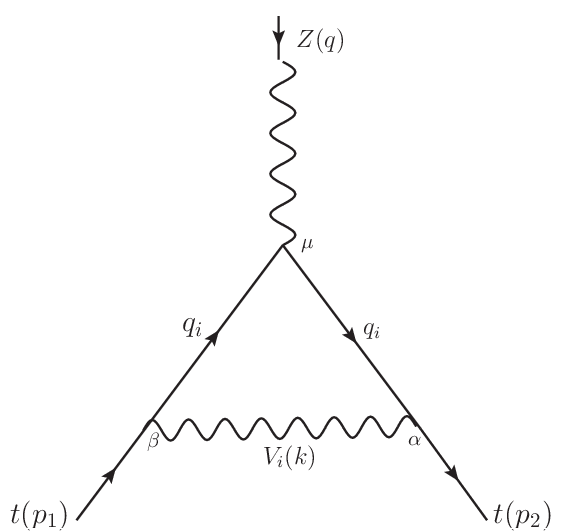}}
\caption{ \label{dipolo} Feynman diagrams contributing to the AWMDM of the top-quark at one-loop.
a) Scalar bosons $S_i=h_0, H_0, A_0, \phi^{0}, \eta^{0}, \sigma, H^{\pm}, \phi^{\pm},
\eta^{\pm}$, b) Vector bosons $V_i= \gamma, Z, W, Z', W'$,  and heavy quarks $q_i=b, t, B, T, T_5, T_6, T^{2/3}, T^{5/3}$.}
\end{figure}

\begin{eqnarray}\label{amplitudesSV1}
\mathcal{M}^{\mu}_{t}(S_{i})&=& \int \frac{d^{4}k}{(2\pi)^{4}} \bar{u}(p_{2}) \left(f^{*}_{S_{i}} +f^{*}_{P_{i}} \gamma^{5}\right) \left[i \frac{\not\! k + \not\!p_{2}+m_{q_i}  }{(k+p_{2})^{2}-m^{2}_{q_{i}}} \right] \left( \gamma^{\mu} (F_{V_i}+F_{A_{i}}\gamma^{5}) \right)  \nonumber \\
&\times & \left[i \frac{ \not\! k + \not\!p_{1}+m_{q_i}  }{(k+p_{1})^{2}-m^{2}_{q_{i}}} \right] \left(f_{S_{i}} +f_{P_{i}} \gamma^{5}\right) u(p_{1}) \left(\frac{i}{k^{2}- m^{2}_{S_{i}}} \right), \\
\mathcal{M}^{\mu}_{t}(V_{i}) &=&  \int \frac{d^{4}k}{(2\pi)^{4}} \bar{u}(p_{2}) \gamma^{\alpha} \left(f^{*}_{V_{i}} +f^{*}_{A_{i}} \gamma^{5}\right) \left[i \frac{\not\! k + \not\! p_{2}+m_{q_i}  }{(k+p_{2})^{2}-m^{2}_{q_{i}}} \right]  \left( \gamma^{\mu} (F_{V_{i}}+F_{A_{i}}\gamma^{5}) \right) \nonumber \\
&\times & \left[i \frac{ \not\! k + \not\! p_{1}+m_{q_i}  }{(k+p_{1})^{2}-m^{2}_{q_{i}}} \right] \gamma^{\beta}\left(f_{V_{i}} +f_{A_{i}} \gamma^{5}\right) u(p_{1})  \left[\frac{i}{k^{2}- m^{2}_{V_{i}}} \left(-g_{\alpha \beta}+  \frac{k_{\alpha}k_{\beta} }{m^{2}_{V_{i}}}\right) \right],\label{amplitudesSV2} 
\end{eqnarray}

\noindent where $f_{S_{i}}, f_{P_{i}}, f_{V_{i}}$ and $ f_{A_{i}}$ denote the form factors of the scalars,
pseudoscalar, vector and axial-vector.  
For the virtual photon case, the longitudinal term of the propagator in Eq.~(\ref{amplitudesSV2}) is absent. For each amplitude we have to pick up only the coefficients of $\sigma^{\mu \nu}
q_{\nu}$ and $\sigma^{\mu \nu} q_{\nu} \gamma^{5}$ shown in Eq.~(\ref{verticeZtt}).
Therefore, the contributions of the new physics to the AWMDM and WEDM of the top-quark are given as follows

\begin{eqnarray}\label{awt}
a^{W}_{t}\equiv [a^{W}_{t}]^{BLHM} &=& [a^{W}_{t}]^{S_i} + [a^{W}_{t}]^{V_i}, \\
d^{W}_{t}\equiv [d^{W}_{t}]^{BLHM} &=&[d^{W}_{t}]^{S_i} + [d^{W}_{t}]^{V_i}.
\end{eqnarray}

In the context of the BLHM, $d^{W}_{t}$ is absent so it does not receive radiative corrections at the one-loop level.

\subsection{Parameters space}

We consider the following input parameters: $m_{A_{0}}$, $m_{\eta_{0}}$ y  $\tan \beta$. The mass of the
pseudoscalar $A_{0}$, which is fixed around 1000 GeV, is in strict agreement with the most recent experimental data on searches
for new scalar particles~\cite{ATLAS:2020gxx}. On the  other hand, the free parameters $m_{4, 5, 6}$~\cite{JHEP09-2010} are
introduced to break all the axial symmetries in the Higgs potential, giving positive masses to all scalars. Specifically,
the $\eta_{0}$ scalar receives a mass equal to $m_{4}=m_{\eta_{0}}=100$ GeV, according to the BLHM, and the restriction
$m_{4}\gtrsim 10$ GeV must be considered~\cite{JHEP09-2010}. The following theoretical constraints on the BLHM parameters are
 imposed, primarily due to perturbativity requirements, such as the value of the mixing angle $\beta$, which is
restricted to be

\begin{eqnarray}\label{cotabeta}
 1 < \text{tan}\ \beta < \sqrt{ \frac{2+2 \sqrt{\big(1-\frac{m^{2}_{h_0} }{m^{2}_{A_0}} \big) \big(1-\frac{m^{2}_{h_0} }{4 \pi v^{2}}\big) } }{ \frac{m^{2}_{h_0}}{m^{2}_{A_0}} \big(1+ \frac{m^{2}_{A_0}- m^{2}_{h_0}}{4 \pi v^{2}}  \big) } -1 }.
\end{eqnarray}

\noindent From this inequality we can extract values for the $\tan \beta$ parameter. In particular, for $m_{A_{0}}=1000$ GeV,
it is obtained that $1 < \tan \beta < 10.45$. For our analysis we choose $\tan \beta=3$,
$m_{A_{0}}=1000$ GeV, $m_{\eta_{0}}=100$ GeV and $F= 4000$ GeV. Due to the characteristics of the BLHM, avoiding fine-tuning requires
light exotic quarks whereas precision electroweak constraints require new heavy gauge bosons~\cite{JHEP09-2010}. In this way, the energy
scale $F$  was chosen to be large enough to ensure that the new gauge bosons are much heavier than the exotic quarks. Recall that the new
heavy gauge bosons develop masses proportional to combination of $f$ and $F$ (see Eqs.~(\ref{mzprima}) and~(\ref{mwprima})).

\begin{figure}[H]
\subfloat[$y_{2} < y_{3}$]{\includegraphics[width=8.0cm]{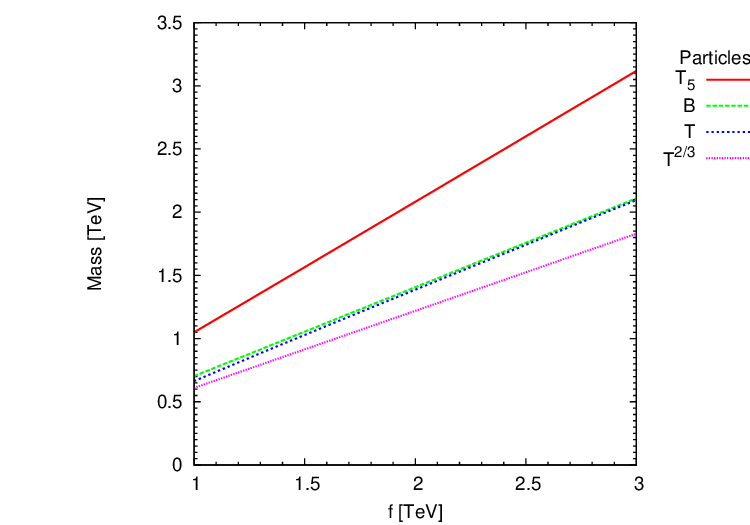}}
\subfloat[$y_{2} > y_{3}$]{\includegraphics[width=8.0cm]{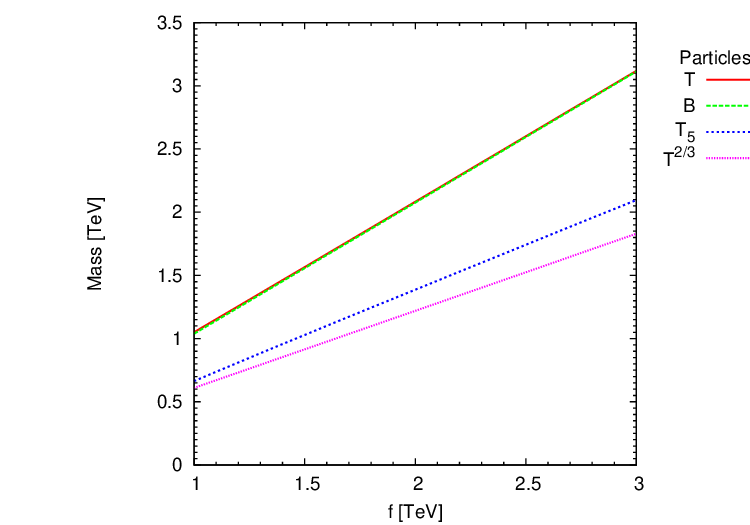}}\\
\centerline{
\subfloat[]{\includegraphics[width=8.0cm]{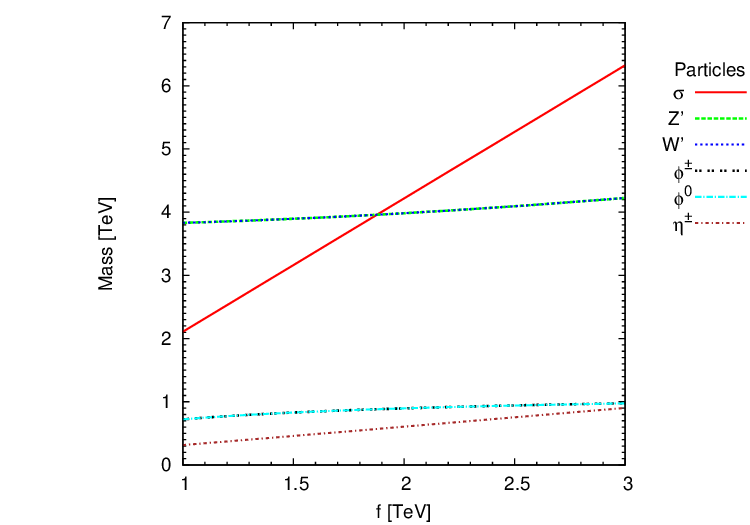}} }
\caption{ \label{massSVQ} Behavior of particle masses as a function of the energy scale $f$ in the BLHM. a) Heavy quarks in the region $y_{2} < y_{3}$. b) Heavy quarks in the region $y_{2} > y_{3}$.
c) Scalars and vector bosons.}
\end{figure}

In Fig.~\ref{massSVQ}, we show the spectrum of the new particles whose masses are proportional to the energy scale $f$. As mentioned in the Subsection~\ref{subsecfermion}, the analytical expressions for the mass eigenvalues, Eqs.~(\ref{mt})~-~(\ref{mB}), are valid only for the region $|y_{2}-y_{3}|>0$. In this way,  the exotic quarks obey the following mass hierarchy in the corresponding sub-regions~\cite{PhenomenologyBLH,Godfrey:2012tf},

\begin{eqnarray}
y_{2} < y_{3}, \ m_{T^{2/3}} &=& m_{T^{5/3}}= m_{T_{6}} < m_{T} < m_{B} < m_{T_5},
\end{eqnarray}

\begin{eqnarray}
y_{2} > y_{3}, m_{T^{2/3}} &=& m_{T^{5/3}}= m_{T_{6}}< m_{T_5} < m_{B}  < m_{T} .
\end{eqnarray}

\noindent When $y_{2} < y_{3}$, the mass difference between the $T_{5}$ and $T_{6}$ 
quarks is large, this increases the decay modes available for the $T_{5}$ state through  cascades decays to non-SM particles.  While for $y_{2} > y_{3}$, the mass splitting between $T_{5}$ and $T_{6}$ is relatively small, so $T_{5}$ state decays predominantly to SM particles~\cite{PhenomenologyBLH,Godfrey:2012tf}. Since the range of parameter space for this model is large, we restrict our study to two sample sub-regions of parameter space that characterize the range of mass for the new heavy quarks, for this purpose we choose: $y_{1}=0.61$, $y_2=0.35$, $y_3=0.84$ ($y_2 < y_3$) and  $y_{1}=0.61$, $y_2=0.84 $, $y_3=0.35$ ( $y_{2}>y_{3}$). The values of the three Yukawa couplings are fixed through the top-quark mass (see Eq.~(\ref{yt})) and satisfy a perturbative scenario. Therefore, with these values of $y_{1,2,3}$ the masses for the heavy quarks are generated, these are provided in Eqs.~(\ref{mT1})~-~(\ref{mB2}). 
In the established scenarios, the lightest quarks are $T^{2/3}$, $T^{5/3}$ and $T_{6}$ as they acquire lower limits on their masses up to 610 GeV~\cite{CMS:2018ubm,CMS:2013wkd,CDF:2009gat,Contino:2008hi}.

\begin{itemize}
\item \text{Prediction with} $y_{2} < y_{3}$:

\begin{eqnarray}
m_{T} &=& [663.3, 2096.8]\  \text{GeV},\label{mT1} \\
m_{T_{5}} &=& [1050.1, 3118.4]\  \text{GeV},\label{mT51}  \\
m_{T^{2/3}} &=& m_{T^{5/3}}= m_{T_{6}} = [610.0, 1830.0]\  \text{GeV}, \label{mT231} \\
m_{B} &=& [703.3, 2109.8]\  \text{GeV}.\label{mB1}
\end{eqnarray}

\item \text{Prediction with} $y_{2} > y_{3}$:
\begin{eqnarray}
m_{T} &=& [1050.1, 3118.4]\  \text{GeV}, \label{mT2} \\
m_{T_{5}} &=& [663.3, 2096.8]\  \text{GeV},\label{mT52}  \\
m_{T^{2/3}} &=& m_{T^{5/3}}= m_{T_{6}} = [610.0, 1830.0]\  \text{GeV}, \label{mT232} \\
m_{B} &=& [1038.1, 3114.4]\  \text{GeV}.\label{mB2}
\end{eqnarray}

\end{itemize}

\noindent With respect to the masses of the scalar and vector particles, in Fig.~\ref{massSVQ} (c) we observe that $\sigma$ is the heaviest scalar, while $\eta^{\pm}$ is the lightest one.
The new gauge bosons $Z'$ and $W'$ acquire  equal masses. Due to the fine-tuning constraints, the scale  $f$ is varied from 1000 to 3000 GeV, thus obtaining a range of masses for each of the new quarks, scalar bosons, and vector bosons.

\begin{eqnarray}
m_{\sigma} &=& [2108.7, 6326.0]\  \text{GeV},\\
%m_{H^{\pm}} &=& m_{A_{0}}, \\
%m_{H_{0}} &\approx & 1010\  \text{GeV} \ \ \ \ (\text{for}\  m_{A_{0}}=1000\ \text{GeV}),\\
m_{\phi^{0}} &=& [722.3, 975.7]\  \text{GeV},\\
m_{\eta^{\pm}} &=& [315.3, 902.7]\  \text{GeV},\\
m_{\phi^{\pm}} &=& [722.5, 976.7]\  \text{GeV}, \\
m_{Z'} &=& m_{W'}=[3831.7, 4225.9]\  \text{GeV}.
\end{eqnarray}

The masses of the scalars $H^{\pm}$ and $H_0$ do not depend on the energy scale $f$  but are calculated from the input parameters of the BLHM,  $m_{A_{0}}$ and $\tan \beta$ (see Eqs.~(\ref{mHmas}) and~(\ref{mH0})).

\begin{eqnarray}
m_{H^{\pm}} &=& m_{A_{0}}, \\
m_{H_{0}} &\approx & 1010\  \text{GeV}, \ \ \ \ \text{for}\  m_{A_{0}}=1000\ \text{GeV}.\label{mH00}
\end{eqnarray}

\subsection{Constraints on model parameters}

The structure of the BLHM  was constructed to solve some problems that occur in most Little Higgs models.
The reason this model succeeds is that it is built under two separate symmetry-breaking scales, $f$ and $F>f$, at which
the exotic quarks and heavy gauge bosons, respectively, obtain their masses. Thus, the gauge bosons are relatively heavy,
consistent with electroweak precision measurements because masses above the already excluded mass range are generated.
For the fermion sector of the model, the most stringent theoretical constraint on the masses of the exotic quarks comes from
fine-tuning of the Higgs potential due to fermion loops. It is therefore important to determine realistic values of
the three Yukawa couplings, $y_{1,2,3}$, and the top-quark Yukawa coupling, $y_{t}$, that evade the fine-tuning
constraints. In this sense, a fit on the Yukawa coupling parameters is required. In the BLHM, the size of the
fine-tuning can be computed in the following way~\cite{JHEP09-2010,PhenomenologyBLH}

\begin{eqnarray} \label{fine-tuning}
\Psi=\frac{| \delta m^{2}_{1} |}{\lambda_0 v^{2} \cos^{2} \beta}, \  \delta m^{2}_{1} = -\frac{27 f^2}{8 \pi^{2} } \frac{ y^{2}_{1} y^{2}_{2} y^{2}_{3}  }{y^{2}_{2}-y^{2}_{3} }\, \text{log} \left( \frac{y^{2}_{1} + y^{2}_{2}}{y^{2}_{1} + y^{2}_{3}} \right).
\end{eqnarray}

\noindent  If $\Psi\sim 1$, this indicates that there is no fine-tuning in the model. On the other hand, the top-quark Yukawa
coupling is determined by

\begin{eqnarray}
y_{t}=\frac{m_{t}}{v \sin \beta}=\frac{3 y_1 y_2 y_3}{\sqrt{\left(y^{2}_{1}+ y^{2}_{2} \right) \left(y^{2}_{1} + y^{2}_{3}  \right)   }},
\end{eqnarray}

\noindent where $m_{t}=172.76$ GeV is the top-quark mass, thus finding $y_{t}=0.74$. With this fixed value of $y_t$,
we can randomize perturbative values for the $y_{1,2,3}$ parameters. In order to obtain a numerical estimate of the fine-tuning
and an upper limit for the  $f$ scale where the new physics does not significantly require fine-tuning, we choose the following
values: $y_1=0.61$, $y_2=0.35$ and $y_3=0.84$, with $f<3100$ GeV.
In this scenario we will carry out our analysis of the AWMDM of the top-quark, as for the $y_2>y_3$ scenario, this could be the
subject of another study shortly soon~\cite{EA}.
Finally, the gauge couplings $g_{A}$ and  $g_{B}$, associated with the $SU(2)_{LA}$ and $SU(2)_{LB}$ gauge bosons,
can be parametrized in a more phenomenological form in terms of a mixing angle $\theta_{g}$ and the  $SU(2)_{L}$ gauge coupling:
$\tan \theta_{g}=g_{A}/g_{B}$  and $g=g_{A} g_{B}/\sqrt{g^{2}_{A}+ g^{2}_{B}} $. For simplicity, we can assume that $\tan \theta_{g}=1$,
which implies that the gauge couplings  $g_{A}$ and  $g_{B}$ are equal. The $g_{A,B}$ values are generated using the restriction $g=0.6525$.

\subsection{Feynman rules}

In order to facilitate the phenomenological study of the weak dipole moments of the top-quark in the BLHM. We provide in Appendix A all the Feynman rules of the interaction vertices obtained in the unitary gauge. These three-point vertices refer to the couplings between gauge bosons and fermions, and scalars and fermions. The complete set of Feynman rules presented in this study was determined using perturbation theory and expanded up to $\mathcal{O}(\frac{1}{f^{2}})$.

\subsection{ The top-quark at the ILC}

Top-quark production in the process $e^{+} e^{-} \rightarrow Z^{*}/\gamma \rightarrow t\bar{t}$  at the International
Linear Collider (ILC)~\cite{Behnke:2013xla,Baer:2013cma,Adolphsen:2013kya} is a powerful tool to determine indirectly the scale of new
physics. Such a machine offers several advantages over a hadron collider such as the LHC, especially in performing SM precision measurementes~\cite{Baer:2013cma} since it provides  an experimentally clean environment without hadronic activity in the state
initial, and the collision energy is accurately  known. The ILC is designed to operate in phase II at a center-of-mass energy of
$\sqrt{s}=500$ GeV, at this energy top-quark pairs are produced numerously well above threshold~\cite{Cao:2015qta}.

Thus, in order to give numerical results to $a^{W}_{t}$, we adopt the same collider parameters
of the $e^{+} e^{-}$ linear collider, that is,  $\sqrt{s}=\sqrt{q^{2}} =500$ GeV. Therefore, we have computed the contributions to the
AWMDM $a^{W}_t$ of the on-shell top-quark with the $Z$ boson at the center-of-mass energy expected for the ILC. On the other hand, in
this same scenario, it was obtained that the WEDM $d^{W}_t $ does not receive contributions to one-loop. In the SM, $d^{W}_t $ only
receives  contributions at three-loops~\cite{Hollik:1998vz,Czarnecki:1996rx}.

\subsection{Numerical results}

To solve the integrals involved in the generic amplitudes, the Passarino-Veltman reduction scheme was implemented in the environment of the Mathematica Feyncalc~\cite{Mertig:1990an} and Package-X~\cite{Patel:2015tea}.  The kinematic conditions were used in these packages, as well as the Gordon identity to eliminate the terms proportional to ($p_1 + p_2 $)$^{\mu}$. After this, the AWMDM of the top-quark is obtained through the relation $a^{W}_{t}= -2 m_t F_{M}(q^{2 })$.  In this study, we do not report the analytical expressions  for $a^{W}_{t}$ because are very large expressions, for this reason, we only report our numerical results.

\begin{figure}[H]
\subfloat[]{\includegraphics[width=8.0cm]{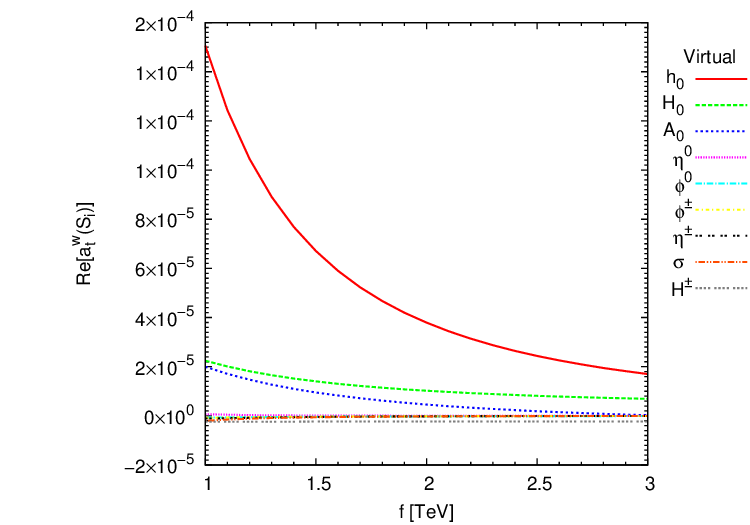}}
\subfloat[]{\includegraphics[width=8.0cm]{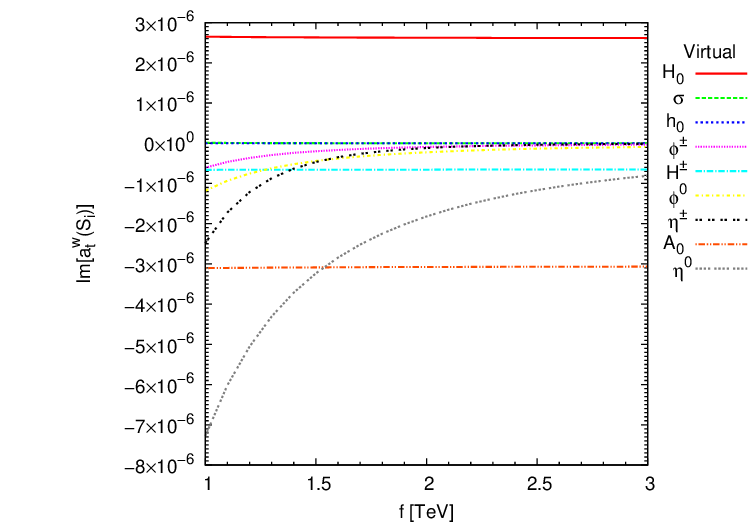}}
\caption{ \label{Si} Individual scalar contributions to $a^{W}_{t}$ in the BLHM for fixed values of $m_{A_{0}}=1000\,
\text{GeV}$, $m_{\eta^{0}}=100\, \text{GeV}$ and $F=4000\, \text{GeV}$.  a) Re($a^{W}_{t}$). b) Im($a^{W}_{t}$).}
\end{figure}

\begin{figure}[H]
\subfloat[]{\includegraphics[width=8.0cm]{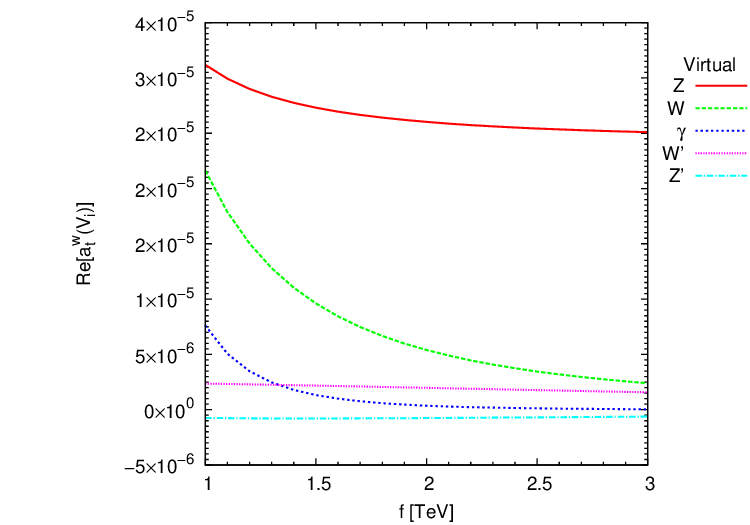}}
\subfloat[]{\includegraphics[width=8.0cm]{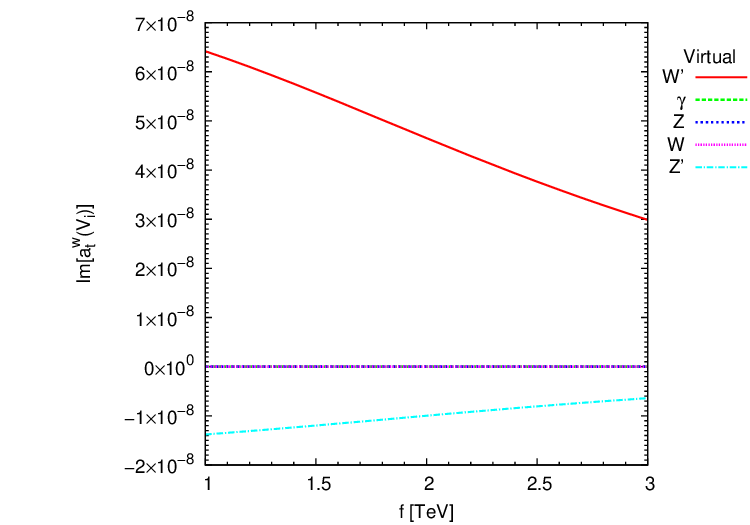}}
\caption{ \label{Vi} Individual vector contributions to $a^{W}_{t}$ in the BLHM for a fixed value of $F=4000\, \text{GeV}$. a) Re($a^{W}_{t}$). b) Im($a^{W}_{t}$).}
\end{figure}

From the 52 diagrams contributing to the $t\bar{t}Z$  vertex,  we start by extracting the contributions to $a^{W}_{t}$ due to the different scalar bosons and vector bosons. In Figs.~(\ref{Si})~-~(\ref{Vi}) is shown $a^{W}_{t}$ as a function of the new physics scale $f$ for the intervalo $f=[1000,3000]$ GeV.
From Fig.~\ref{Si} we can appreciate the individual contributions of the scalars involved, and we observe that the main contributions to the real  and imaginary part of $a^{W}_{t}$ are generated by the Higgs bosons $h_0$ and $H_0$: $\text{Re}[ a^{W}_{t} (h_{0} ) ] =[1.51 \times 10^{-4}, 1.70 \times 10^{-5}]$ and $\text{Im} [ a^{W}_{t} (H_{0})] =[2.65,2.62 ]\times 10^{-6}$.
In contrast, the smallest contributions are provided by the charged scalar $H^{\pm}$ and the neutral scalar $\eta^{0}$: $\text{Re}[ a^{W}_{t} (H^{\pm} )  ] = - [2.38, 2.33] \times 10^{-6}$ and $\text{Im}[ a^{W}_{t}(\eta^{0} ) ] = -[7.28 \times 10^{-6}, 8.09 \times 10^{-7}] $. With respect to the remaining scalars, these are suppressed by one or up to three orders of magnitude compared to the absolute value of the main or smallest contributions in their class.
In Fig.~\ref{Vi}, all the contributions to the AWMDM of the top-quark, coming from virtual vector particles
are displayed. In this figure, the dominant contributions to $a^{W}_{t}$ are generated when the intermediate
particles are the $Z$ and $W'$ gauge bosons, that is, $ \text{Re}[ a^{W}_{t} (Z)  ] = [3.12,2.51] \times 10^{-5} $ and  $ \text{Im}[ a^{W}_{t} (W' ) ] = [6.42,2.99]\times 10^{-8}$. The minor contributions acquire negative values and  occur for $Z'$ bosons: $ \text{Re}[ a^{W}_{t}(Z' ) ] = - [7.39, 6.22]\times 10^{-7} $ and $ \text{Im}[ a^{W}_{t} (Z' ) ] = - [1.38 \times 10^{-8},6.40 \times 10^{-9}] $. For the real part of $a^{W}_{t}$, the virtual bosons $W$, $\gamma$ and $W'$ contribute on the order of $10^{-5}$ to $10^{-8}$.  While for the imaginary part, the numerical contributions of the vector bosons $\gamma$, $Z$ and $W$ are zero.
Note here that the mediator particles $h_0$ and $Z$  are all from the SM and these provide the largest positive contributions to $ \text{Re}[ a^{W}_{t}]$, on the other hand, for $ \text{Im}[ a^{W}_{t}]$ the new exotic particles $H_0$ and $W'$ are the ones that contribute significantly more than the others.
In Appendix B,
%(see Tables~\ref{CN1}~-~\ref{CN2})
we present all BLHM contributions to the AWMDM in the $t\bar{t}Z$ vertex.

\begin{figure}[H]
\subfloat[]{\includegraphics[width=8.0cm]{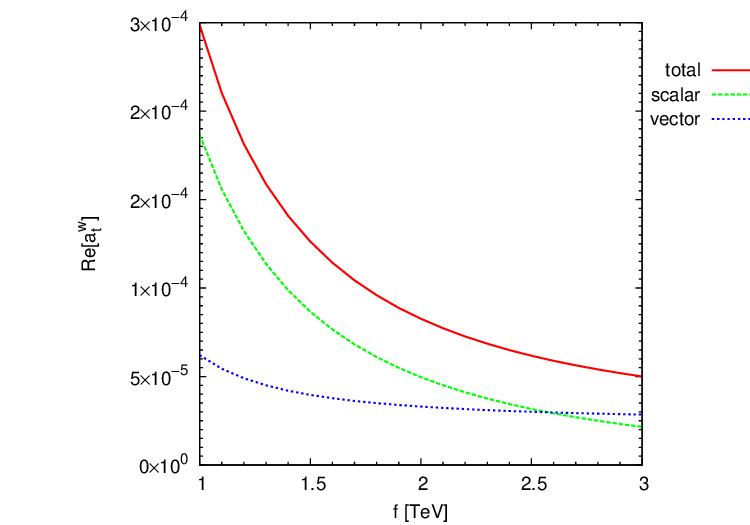}}
\subfloat[]{\includegraphics[width=8.0cm]{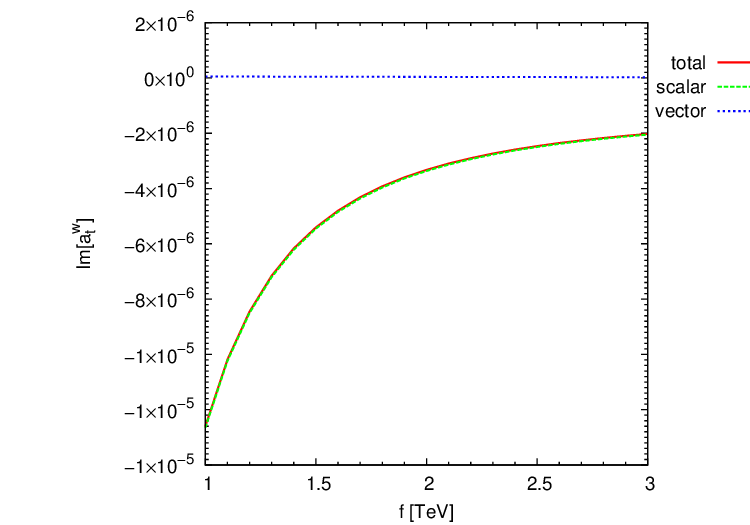}}
\caption{ \label{FS} Scalar, vector and total contributions  to $a^{W}_{t}$ in the BLHM for fixed values of $m_{A_{0}}=1000\,
\text{GeV}$, $m_{\eta^{0}}=100\, \text{GeV}$ and $F=4000\, \text{GeV}$. a) Re($a^{W}_{t}$). b) Im($a^{W}_{t}$).}
\end{figure}

The total contribution on the AWMDM of the top-quark involving scalar bosons, vector bosons, and exotic quarks in the loop is given in Fig.~\ref{FS}. 
In the range of analysis established for the $f$ scale, the total contribution on the AWMDM receives contributions coming mainly from two sectors: scalar and vector.
Each sector arises from the sum of all scalar and vector contributions, respectively. In this manner, in  Fig.~\ref{FS} (a) we can appreciate the real contributions to $a^{W}_{t}$ and we find that the significant contribution to the total contribution comes from the scalar contribution, as both of which contribute
in the order of magnitude of $10^{-4}$ to $10^{-5}$. On the other hand, the subdominant contribution is generated by the vectors:
 $\text{Re}[a^{W}_{t}(\text{vector})]\sim 10^{-5}$ for $f \in [1000,3000]$ GeV. Concerning Fig.~\ref{FS} (b), the relevant contribution to the imaginary part of the AWMDM of the top arises from the scalar contribution, this contributes to the order of  $10^{-5}$ to $10^{-6}$. The vector contribution is of the order of $10^{-8}$. The values of the total contribution to  $a^{W}_{t}$ are listed in Table~\ref{C1000}.

\begin{figure}[H]
\subfloat[]{\includegraphics[width=8.0cm]{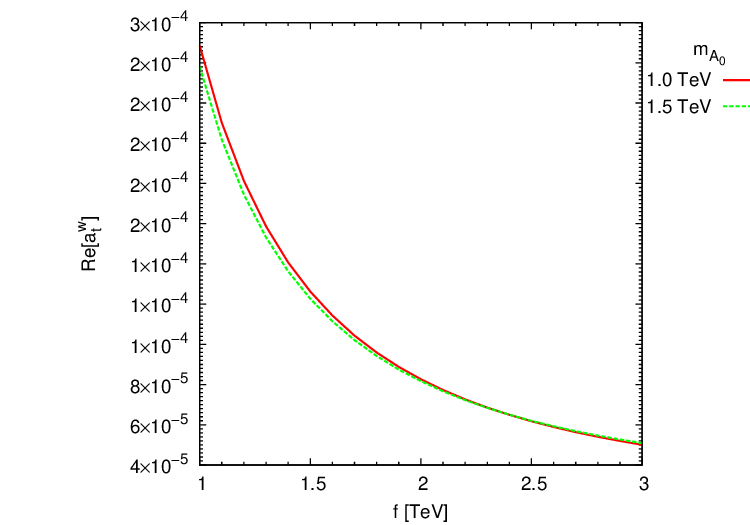}}
\subfloat[]{\includegraphics[width=8.0cm]{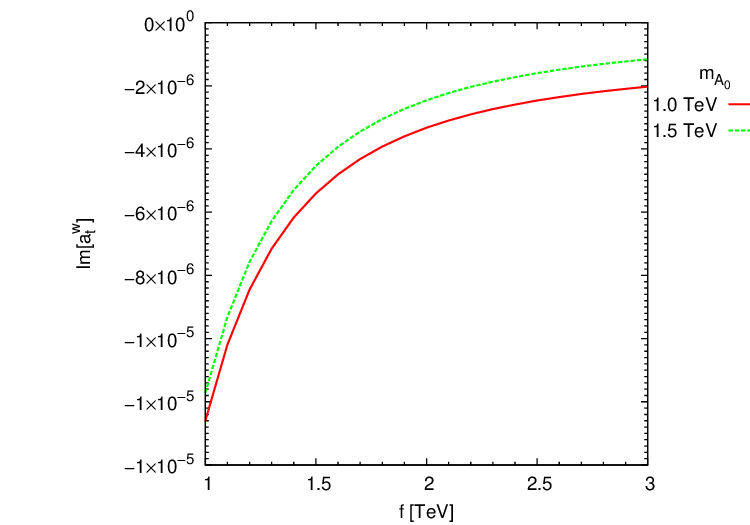}}
\caption{ \label{FmA} Total contribution to $a^{W}_{t}$ in the BLHM for different values of the mass of $A_0$ but fixed
values of  $m_{\eta^{0}}=100\, \text{GeV}$ and $F=4000\, \text{GeV}$. a) Re($a^{W}_{t}$). b) Im($a^{W}_{t}$).}
\end{figure}

\begin{figure}[H]
\subfloat[]{\includegraphics[width=8.0cm]{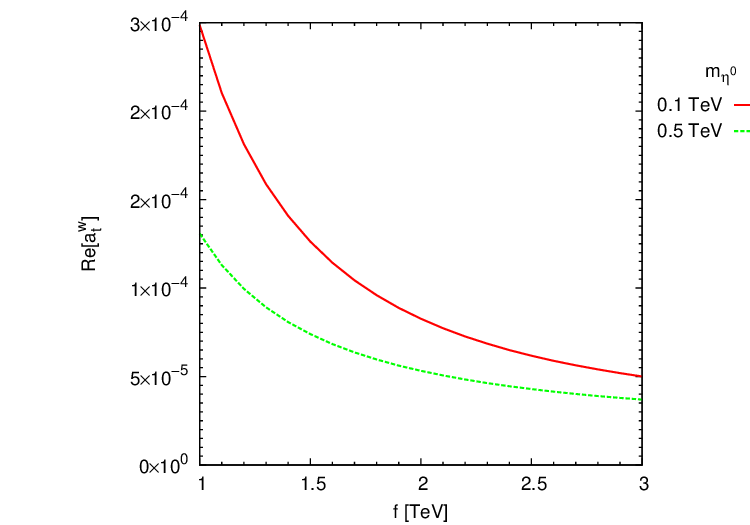}}
\subfloat[]{\includegraphics[width=8.0cm]{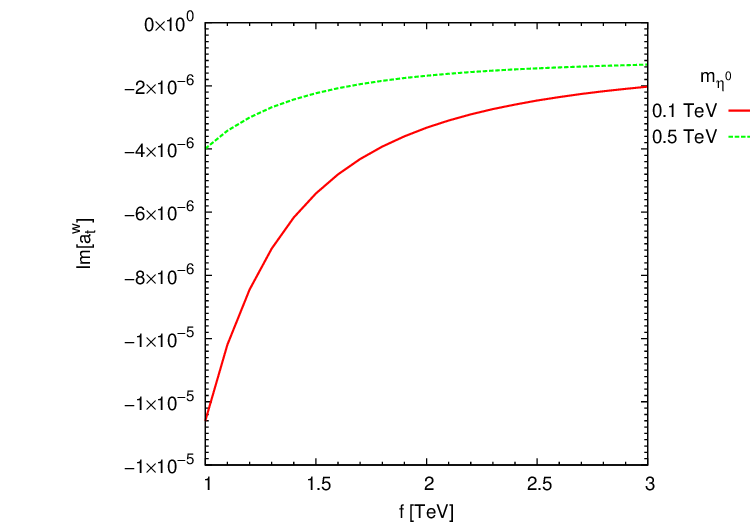}}
\caption{ \label{Fmeta} Total contribution to $a^{W}_{t}$ in the BLHM for different values of the mass of $\eta^{0}$
and fixed values of $m_{A_{0}}=1000\, \text{GeV}$ and $F=4000\, \text{GeV}$. a) Re($a^{W}_{t}$). b) Im($a^{W}_{t}$).}
\end{figure}

Since the masses of the states, $A_{0}$ and $\eta_{0}$ are input parameters of the BLHM,
in order to measure the sensitivity of the AWMDM of the top-quark, we allow us to vary the masses of the mentioned states.
For the pseudoscalar $A_{0}$ we have chosen the values of $m_{A_{0}}=1000, 1500$ GeV, while for the neutral scalar $\eta^{0}$
we chose $m_{\eta^{0}}=100, 500$ GeV. In this way, we generate the Figs. \ref{FmA} and \ref{Fmeta} that show the behavior of the real and imaginary
part of $a^{W}_{t}$ when the energy scale $f$ varies from $1000$ GeV to $3000$ GeV, while fixing the other free parameters of the model. With the two fixed values of $m_{A_{0}}$,
identical plots are generated as shown in Fig. \ref{FmA} (a), this indicates that Re[$a^{W}_{t}$] is indifferent to the chosen values of the mass of the  pseudoscalar $A_0$.
Fig. \ref{FmA} (b) shows  plots with a very slight difference, since both contribute to the same order of magnitude. 
For the established values of $m_{\eta^{0}}$, in Fig. \ref{Fmeta} (a) we observe that
$|\text{Re}\, [a^{W}_{t}]|$ obtains more intense values when the mass of the $\eta^{0}$ scalar is small, specifically when
$m_{\eta^{0}}=100$ GeV, $|\text{Re}\, [a^{W}_{t}]| \in [2.49 \times 10^{-4}, 5.01 \times 10^{-5}]$. 
The same pattern occurs in Fig. \ref{Fmeta} (b), in this case $|\text{Im}\, [a^{W}_{t}]|\in [1.26 \times 10^{-5}, 2.03 \times 10^{-6}]$  for $m_{\eta^{0}}=100$ GeV.
This result is to be expected since the contribution to AWMDM of the top-quark decouples as $m_{\eta^{0}}$ increases.
For the different scenarios considered above, we provide the values of $a^{W}_{t}$ in Tables \ref{C1000}-\ref{C1500}.

\begin{figure}[H]
\subfloat[]{\includegraphics[width=8.0cm]{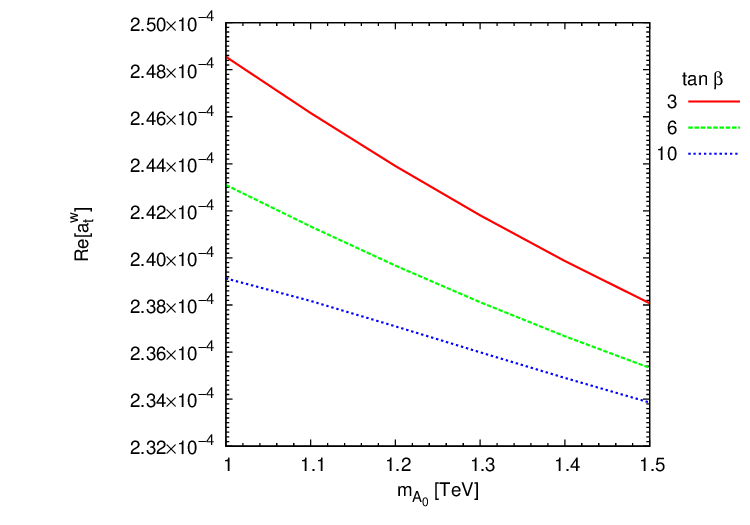}}
\subfloat[]{\includegraphics[width=8.0cm]{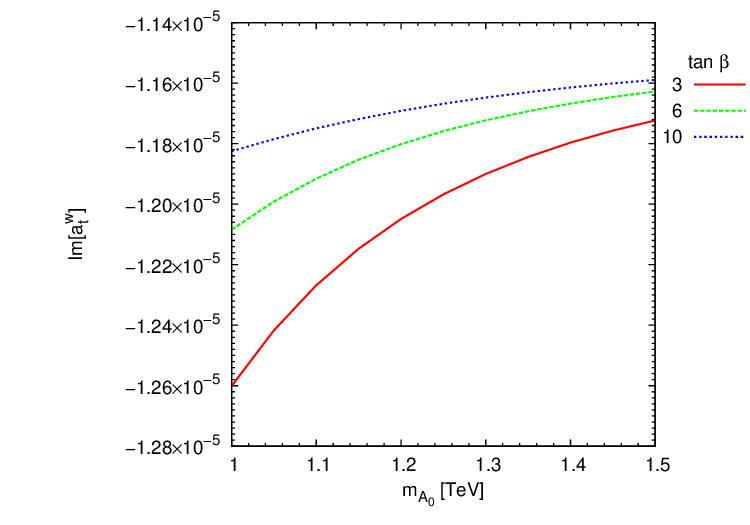}}
\caption{ \label{Ftan} Total contribution to $a^{W}_{t}$ in the BLHM for different values of $\tan \beta$
and fixed values of $f=1000\, \text{GeV}$, $m_{\eta^{0}}=100\, \text{GeV}$ and $F=4000\, \text{GeV}$.
a) Re($a^{W}_{t}$). b) Im($a^{W}_{t}$).}
\end{figure}

Another input parameter of the BLHM is $\tan \beta$, the values it acquires are restricted to the intervals generated according to  Eq. (\ref{cotabeta}). Therefore, if
$m_{A_{0}}$ varies from 1000 GeV to 1500 GeV, $\tan \beta \in  (1,10.45)$. For certain fixed values of $\tan \beta$, we obtain
the behavior of $a^{W}_{t}$ as a function of the mass of the pseudoscalar $A_0$. 
In Fig. \ref{Ftan}, we observe that $|\text{Re}\, [a^{W}_{t}]|$ obtains large values when $\tan \beta=3$ while $\tan \beta=6$ and $\tan \beta=10$ yield suppressed contributions to $a^{W}_{t}$.
With respect to $|\text{Im}\, [a^{W}_{t}]|$, it also acquires high values when $\tan \beta=3$.
According to Figs.~\ref{Ftan} (a)~and~\ref{Ftan} (b) , it can be seen that for the different fixed values of $\tan \beta$,  $|\text{Re}\, [a^{W}_{t}]|\sim 10^{-4}$ and $|\text{Im}\, [a^{W}_{t}]| \sim 10^{-5}$ are obtained.

\begin{figure}[H]
\subfloat[]{\includegraphics[width=8.0cm]{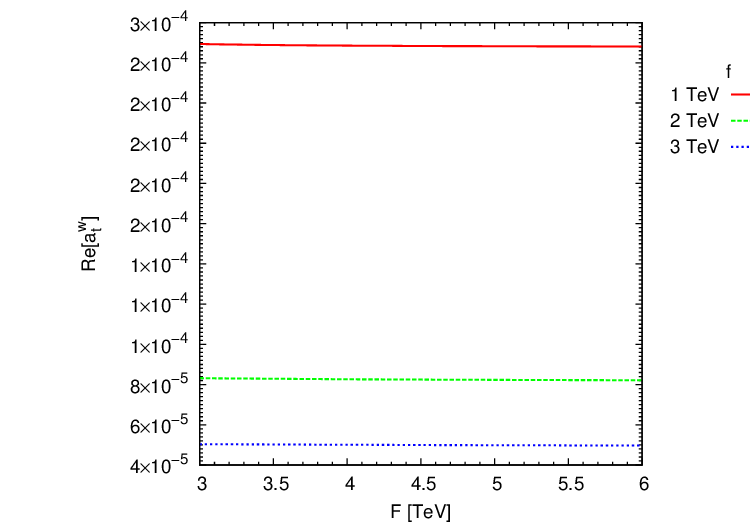}}
\subfloat[]{\includegraphics[width=8.0cm]{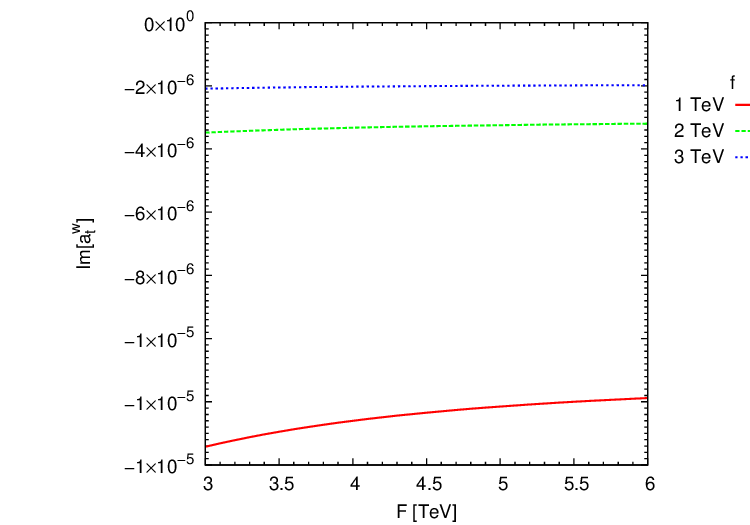}}
\caption{ \label{Ffi} Total contribution to $a^{W}_{t}$ in the BLHM for different values of the energy scale $f$ with
fixed values of $m_{A_{0}}=1000\, \text{GeV}$ and $m_{\eta^{0}}=100\, \text{GeV}$. a) Re($a^{W}_{t}$). b) Im($a^{W}_{t}$).}
\end{figure}

\begin{figure}[H]
\subfloat[]{\includegraphics[width=8.0cm]{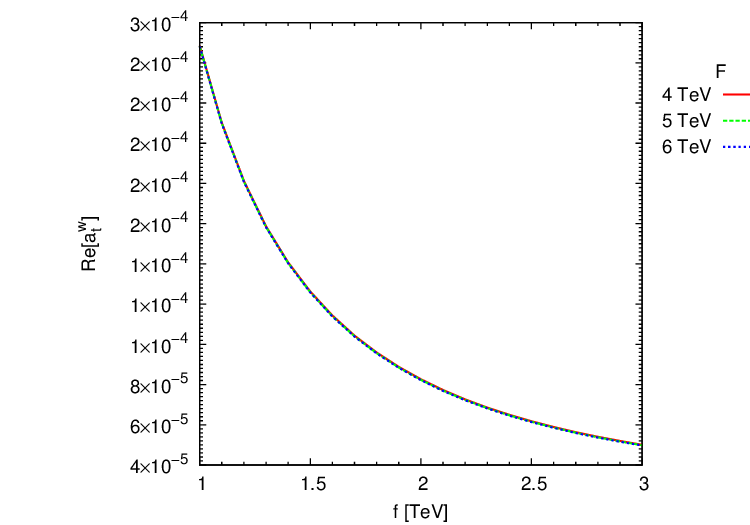}}
\subfloat[]{\includegraphics[width=8.0cm]{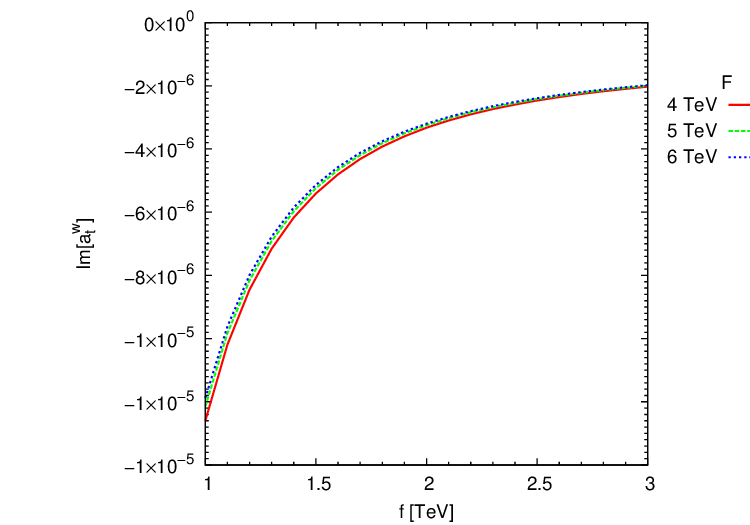}}
\caption{ \label{FFi} Total contribution to $a^{W}_{t}$ in the BLHM for different values of the energy scale $F$ with
fixed values of $m_{A_{0}}=1000\, \text{GeV}$ and $m_{\eta^{0}}=100\, \text{GeV}$. a) Re($a^{W}_{t}$). b) Im($a^{W}_{t}$).}
\end{figure}

The two symmetry breaking energy scales are also free and important parameters of the BLHM,
which we can vary. First, it can be seen from Fig. \ref{Ffi} (a) that $|\text{Re}\, [a^{W}_{t}]|$ does not essentially
depend on $F$ parameter variations, while in Fig. \ref{Ffi} (b), $|\text{Im}\, [a^{W}_{t}]|$ does depend slightly on $F$.
These plots show a variation of the $F$ parameter from $3000$ GeV to $6000$ GeV, for three distinct energy scales, i.e.
$f= 1, 2, 3$ TeV. In these figures, the main contributions to $|a^{W}_{t}|$ are generated for $f= 1$ TeV,  while secondary
contributions arise for $f= 2$ TeV. Second, Fig. \ref{FFi} visualizes the behavior of $|\text{Re}\, [a^{W}_{t}]|$ and
$|\text{Im}\, [a^{W}_{t}]|$ as a function of the $f$ scale. In this case, the graphs do depend on the variations of the
parameter $f$. We have adopted the followings specific values for the parameter $F$, $F= 4, 5, 6$ TeV so that the contribution
to $|a^{W}_{t}]|$ is enhanced. However, there is not significant difference in the behavior of these plots. According to
Eq.~(\ref{fine-tuning}), the energy scale $f$ is also intimately related to the measure of the fine-tuning, and we observed
that for $f=1$ TeV not only large contributions are generated for $|[a^{W}_{t}]|$, we also ensure the absence of fine-tuning.

We have calculated the one-loop level contributions on the AWMDM of the top-quark in several scenarios.
We found that the real and imaginary parts of $a^W_t$ is in the range $10^{-4}$ to $10^{-12}$ when the parameters $f$,
$F$ and $m_{A_{0}}$ are varied in the corresponding intervals: $f=[1000,3000]$ GeV, $F=[3000,6000]$ GeV and $m_{A_{0}}=[1000,1500]$ GeV.
The scalar sector provide
the largest numerical contributions to $|\text{Re}\, [a^{W}_{t}]|$ and $|\text{Im}\, [a^{W}_{t}]|$. 
In general, $|\text{Im}\, [a^{W}_{t}]|$ is  suppressed compared to $|\text{Re}\, [a^{W}_{t}]|$.
In addition, we have found that our results are sensitive to changes in the value of $m_{\eta^{0}}$.
 The contribution to AWMDM from the top-quark decouples with large masses of the $ \eta^{0} $ scalar, as shown in Fig.~\ref{Fmeta}.
Since our main goal in this work is to study the effect of the new particles generated in the BLHM framework, our results reported in this work on $a^{W}_{t}$ depend on the energy scales $f$ and $F$, which represent the scales of the new physics.  In this scenario, we have found that the numerical values found for the AWMDM of the top-quark are comparable with the predictions of the SM or BSM. 
In the context of the SM, Bernabeu et al.~\cite{Bernabeu:1995gs}, they found that the numerical predictions for $a^{W}_{t}$ are on the order of $10^{-3}$.
In Ref.~\cite{Bernreuther:2005gq} the one-loop level QCD contribution to the AWMDM of the top quark in the SM scenario has also been calculated, finding that $a^{W}_{t}=5.2\times 10^{-3}$ for renormalization scale $\mu=m_{t}=175$ GeV.
Within the Minimal Supersymmetric Standard Model (MSSM), the contributions found for the AWMDM of the top-quark are on the order of magnitude of $10^{-4}$~\cite{Bartl:1997iq}.  In models with Two Higgs Doublets (2HDM), the induced effects of new scalars in the $t\bar{t}Z$  vertex loop were also investigated, obtaining $a^{W}_{t} \sim 10^{-3}$~\cite{Bernabeu:1995gs}.
Other studies performed in extended models that predict the existence of new $Z'$ gauge boson, is obtained $a^{W}_{t} \in [10^{-6},10^{-10}]$~\cite{Vivian:2019zfa}.

On the experimental side, the weak dipole moments of the top-quark have not yet been tested directly. There
are promising project at colliders such as the LHC~\cite{ATLAS:2016wgc,Rontsch:2015una}, the future ILC~\cite{Behnke:2013xla,Baer:2013cma,Adolphsen:2013kya}, the CLIC~\cite{deBlas:2018mhx,Robson:2018enq,Roloff:2018dqu,CLIC:2016zwp,Abramowicz:2016zbo} and the Future Circular Collider hadron-hadron (FCC-hh)~\cite{Barletta:2014vea,Koratzinos:2015fya}
that have as an important part of their physics program  to investigate and constrain the  dipole moments of the top-quark, in particular, the AWMDM.
Currently, in the LHC experiment, the most hopeful avenue for studying the top-quark electroweak couplings is through the $pp\rightarrow \bar{t}tV$ ($V=Z,\gamma,H$) processes, which produces direct sensitivity without intrinsic dilution by QCD effects. In the FCC with proton-proton collisions, SM particles will also be produced in great abundance, and in this case, the study of the AWMDM  can be carried out through the $pp\rightarrow \bar{t}tZ$  process. At a leptonic collider such as the ILC or CLIC,  the focus of investigation will be via the $e^{+}e^{-}\rightarrow Z^{*}/ \gamma \rightarrow \bar{t}t $ process which is extremely sensitive to top-quark electroweak couplings.

From the phenomenological point of view, for instance, through the production cross-section of  $t\bar{t}ZZ$, limits on $a^{W}_{t}$ are estimated at $95\ \%$ C.L. corresponding to $3000$ fb$^{-1}$ of integrated luminosity at the LHC: $-0.1 \leq  a^{W}_{t} \leq 0.09$~\cite{Etesami:2017ufk}. In this same integrated luminosity scenario, and through the $p^{Z}_{T}$ distribution in $\bar{t}tZ$ production, it is  found that $ -0.08 \lesssim a^{W}_{t} \lesssim 0.08$~\cite{Rontsch:2015una}.
FCC-hh will provide collisions at a center-of-mass energy of 100 TeV, a factor 7 higher than the LHC. At this energy and with 10 000 fb$^{-1}$ of data, about $10^{8}$ $\bar{t}tZ$ events will be produced. At a rough estimate, the FCC-hh will provide improved limits to $a^{W}_{t}$ by a factor of 3 to 10 compared to the 3000 fb$^{-1}$ LHC~\cite{Rontsch:2015una,Barletta:2014vea,Koratzinos:2015fya}.
Finally, at the ILC with $\sqrt{s}=500$ GeV and 500 fb$^{-1}$, the limits of $-0.02 \lesssim a^{W}_{t} \lesssim 0.04$ at $95 \ \%$ C.L. are expected
to be reached. They are derived by exploiting the total cross-section of the top-quark pair production~\cite{Rontsch:2015una}.
The ILC offers the possibility of extending the LHC top-quark program~\cite{Baer:2013cma} and is one of the most advanced proposals for an $e^{+}e^{-}$ collider.
In the case of the ILC there is an improvement in the $a^{W}_{t}$ by a factor of three (through $\bar{t}tZ$ production) or four (through $\bar{t}tZZ$ production)  compared to the LHC constraints. For these reasons, we believe that our results can be verified by the ILC in the future because it has the potential to reach the required level of sensitivity.

\section{Conclusions}

In this paper, we present a new comprehensive study on the sensitivity  limits of the AWMDM of the top-quark in the context of the BLHM
at the one-loop level. In our study, we have taken into account all contributions from the scalar sector, gauge sector,
and heavy quarks. For which we deduced the allowed ranges of the masses of the new quarks, scalars, and vector bosons (see Eqs.~(\ref{mT1})~-~(\ref{mH00})),
as well as the corresponding Feynman rules of the BLHM expanded up to $\mathcal{O}(\frac{1}{f^{2}})$ (see Tables~\ref{FeyRul-A0}~-~\ref{FeyRul-current-4}). These results are an original contribution.

The sensitivity on the AWMDM of the top-quark has been explored in the region of the parameter space allowed by the fine-tuning
constraints, and measured by varying the main free parameters of the BLHM: $m_{A_{0}}$, $m_{\eta^{0}}$, $f$, $F$ and $\tan \beta$.
Our results are summarized through a set of Figs.~(\ref{massSVQ})~-~(\ref{FFi}), Tables~\ref{C1000}~-~\ref{C1500} (Sensitivity limits
on the AWMDM at the BLHM) and Tables~\ref{FeyRul-A0}~-~\ref{FeyRul-current-4}
(Feynman rules for the BLHM). We find that with the appropriated parameters of the BLHM it is possible to put limits
on the AWMDM of the top-quark with a sensitivity of the order of $a^{W}_{t}= 2.49
\times 10^{-4}-1.26 \times 10^{-5}i$, $1.26 \times 10^{-4}-5.41\times 10^{-6}i$, where the main
contribution comes from  the scalars $h_0$ and $H_{0}$.
The sensitivity limits on $a^W_{t}$ obtained in the context of the BLHM (see Tables \ref{C1000}~-~\ref{C1500})
 are competitive concerning for to the reported in Refs.~\cite{Bernabeu:1995gs,Bernreuther:2005gq,Bartl:1997iq,Vivian:2019zfa}, and in some cases compare favorably.
We should remark that our results found for $a^W_{t}$ fall within the phenomenological bounds provided by colliders such as LHC, FCC-hh and ILC~\cite{Etesami:2017ufk,Rontsch:2015una,Barletta:2014vea,Koratzinos:2015fya}.
Present, there are no precision experimental measurements on the AWMDM of the top-quark. However, future proposed experiments are expected to reach sensitivity to predicted values for observables in the BLHM.
In addition,
as this topic is worthwhile yet underexplored, theoretical, experimental and phenomenological  interest is of great importance in order to motivate
experimental collaborations to measure this very intriguing sector of the SM, which could give evidence of new physics BSM.

\vspace{7cm}

\begin{table}[H]
\caption{Expected sensitivity limits on the $a^W_t$ in the context of the BLHM with $\sqrt{q^{2}}=500\ \text{GeV}$,
\ $\text{m}_{A_0}=1000\ \text{GeV}$, $\text{m}_{\eta_0}=100\ \text{GeV} $, $F=4000\ \text{GeV}$ and $f=1, 1.5, 2, 2.5, 3\ \text{TeV}$
are represented. All new contributions are considered, scalar bosons, vector bosons and heavy quarks.
\label{C1000}}
\centering
\begin{tabular}{|c|c|}
\hline
\hline
\multicolumn{2}{|c|}{$\sqrt{q^{2}}=500\ \text{GeV}$,\ $\text{m}_{A_0}=1000\ \text{GeV}$, $\text{m}_{\eta_0}=\bf{ 100}\ \text{GeV} $,\ $F=4000\ \text{GeV}$}\\
\hline
 $f\ [\text{TeV}]$  & $(a^{W}_{t})^{\rm Total} $  \\
\hline
\hline
$1_\cdot0$  & $2_\cdot 49 \times 10^{-4} - 1_\cdot 26 \times 10^{-5}\ i $  \\
\hline
$1_\cdot5 $  & $ 1_\cdot 26 \times 10^{-4} - 5_\cdot 41 \times 10^{-6}\ i$  \\
\hline
$2_\cdot0 $  & $ 8_\cdot 26 \times 10^{-5} - 3_\cdot 33 \times 10^{-6}\ i$  \\
\hline
$2_\cdot5 $  & $ 6_\cdot 18 \times 10^{-5} - 2_\cdot 47\times 10^{-6}\ i$  \\
\hline
$3_\cdot0 $  & $ 5_\cdot 01 \times 10^{-5} - 2_\cdot 03 \times 10^{-6}\ i$  \\
\hline
\end{tabular}
%\caption{Some values of $a^{W}_{t}$.
%\label{C1000}}
\end{table}

\begin{table}[H]
\caption{Expected sensitivity limits on the $a^W_t$ in the context of the BLHM with $\sqrt{q^{2}}=500\ \text{GeV}$,
\ $\text{m}_{A_0}=1000\ \text{GeV}$, $\text{m}_{\eta_0}=500\ \text{GeV}$, $F=4000\ \text{GeV}$ and $f=1, 1.5, 2, 2.5, 3\ \text{TeV}$
are represented. All new contributions are considered, scalar bosons, vector bosons and heavy quarks.
\label{eta100}}
\centering
\begin{tabular}{|c|c|}
\hline
\hline
\multicolumn{2}{|c|}{$\sqrt{q^{2}}=500\ \text{GeV}$,\ $\text{m}_{A_0}=1000\ \text{GeV}$, $\text{m}_{\eta_0}=\bf{500}\ \text{GeV}$,  $\ F=4000\ \text{GeV}$}\\
\hline
 $f\ [\text{TeV}]$  & $(a^{W}_{t})^{\rm Total} $  \\
\hline
\hline
$1_\cdot0$  & $ 1_\cdot 31 \times 10^{-4} - 3_\cdot 99 \times 10^{-6}\ i$  \\
\hline
$1_\cdot5 $  & $ 7_\cdot 40 \times 10^{-5} - 2_\cdot 23 \times 10^{-6}\ i$  \\
\hline
$2_\cdot0 $  & $ 5_\cdot 32 \times 10^{-5} - 1_\cdot 68 \times 10^{-6}\ i$  \\
\hline
$2_\cdot5 $  & $ 4_\cdot 29 \times 10^{-5} - 1_\cdot 45  \times 10^{-6}\ i$  \\
\hline
$3_\cdot0 $  & $ 3_\cdot 69 \times 10^{-5} - 1_\cdot 33 \times 10^{-6}\ i$  \\
\hline
\end{tabular}
%\caption{Some values of $a^{W}_{t}$.
%\label{eta100}}
\end{table}

%\begin{table}[H]
%\caption{Expected sensitivity limits on the $a^W_t$ in the context of the BLHM with $\sqrt{q^{2}}=500\ \text{GeV}$,
%\ $\text{m}_{A_0}=1000\ \text{GeV}$, $\text{m}_{\eta_0}=1000\ \text{GeV}$, $F=4000\ \text{GeV}$ and $f=1, 1.5, 2, 2.5, 3\ \text{TeV}$
%are represented. All new contributions are considered, scalar bosons, vector bosons and heavy quarks.
%\label{eta1000}}
%\centering
%\begin{tabular}{|c|c|}
%\hline
%\hline
%\multicolumn{2}{|c|}{$\sqrt{q^{2}}=500\ \text{GeV}$,\ $\text{m}_{A_0}=1000\ \text{GeV}$, $\text{m}_{\eta_0}=\bf{1000}\ \text{GeV}$,  $\ F=4000\ \text{GeV}$}\\
%\hline
% $f\ [\text{TeV}]$  & $(a^{W}_{t})^{\rm Total} $  \\
%\hline
%\hline
%$1_\cdot0$  & $ -2_\cdot 36 \times 10^{-3} - 1_\cdot 70 \times 10^{-6}\ i$  \\
%\hline
%$1_\cdot5 $  & $ -1_\cdot 05 \times 10^{-3} - 1_\cdot 28\times 10^{-6}\ i$  \\
%\hline
%$2_\cdot0 $  & $ - 5_\cdot 88 \times 10^{-4} - 1_\cdot 16\times 10^{-6}\ i$  \\
%\hline
%$2_\cdot5 $  & $ -3_\cdot 74 \times 10^{-4} - 1_\cdot 11 \times 10^{-6}\ i$  \\
%\hline
%$3_\cdot0 $  & $ -2_\cdot 59 \times 10^{-4} - 1_\cdot 09 \times 10^{-6}\ i$  \\
%\hline
%\end{tabular}
%%\caption{Some values of $a^{W}_{t}$.
%%\label{eta1000}}
%\end{table}

\begin{table}[H]
\caption{Expected sensitivity limits on the $a^W_t$ in the context of the BLHM with $\sqrt{q^{2}}=500\ \text{GeV}$,
\ $\text{m}_{A_0}=1 500\ \text{GeV}$, $\text{m}_{\eta_0}=100\ \text{GeV}$, $F=4000\ \text{GeV}$ and $f=1, 1.5, 2, 2.5, 3\ \text{TeV}$
are represented. All new contributions are considered, scalar bosons, vector bosons and heavy quarks.
\label{C1500}}
\centering
\begin{tabular}{|c|c|}
\hline
\hline
\multicolumn{2}{|c|}{$\sqrt{q^{2}}=500\ \text{GeV}$,\ $\text{m}_{A_0}=\bf{1500}\ \text{GeV}$, $\text{m}_{\eta_0}=100\ \text{GeV}$,  $\ F=4000\ \text{GeV}$}\\
\hline
 $f\ [\text{TeV}]$  & $(a^{W}_{t})^{\rm Total} $  \\
\hline
\hline
$1_\cdot0$  & $ 2_\cdot 38 \times 10^{-4} -1_\cdot 17 \times 10^{-5}\ i$  \\
\hline
$1_\cdot5 $  & $  1_\cdot 23 \times 10^{-4} - 4_\cdot 53\times 10^{-6}\ i$  \\
\hline
$2_\cdot0 $  & $ 8_\cdot 17 \times 10^{-5} - 2_\cdot 46\times 10^{-6}\ i$  \\
\hline
$2_\cdot5 $  & $ 6_\cdot 20\times 10^{-5} - 1_\cdot 60 \times 10^{-6}\ i$  \\
\hline
$3_\cdot0 $  & $ 5_\cdot 09 \times 10^{-5} - 1_\cdot 16 \times 10^{-6}\ i$  \\
\hline
\end{tabular}
%\caption{Some values of $a^{W}_{t}$.
%\label{C1500}}
\end{table}

%\begin{table}[H]
%\caption{Expected sensitivity limits on the $a^{W}_{t}$ in the decoupling limit between the SM and the BLHM.
%All the contributions are considered scalar bosons, vector bosons, gluon and heavy quarks.
%\label{C-SM}}
%\centering
%\begin{tabular}{|c|c|}
%\hline
%\multicolumn{2}{|c|}{$\sqrt{q^{2}}=500\ \text{GeV}$}\\
%\hline
%%\hline
% $\textbf{Couplings abc}$  & $\left( a^{W}_{t} \right)^{ \textbf{abc} } $  \\
%\hline
%\hline
%$ g tt $  & $ -1_\cdot 04 \times 10^{-3} + 1_\cdot 79 \times 10^{-3} \ i$  \\
%\hline
%$ \gamma tt$  & $ -2_\cdot 15 \times 10^{-5} + 3_\cdot 69 \times 10^{-5} \ i$  \\
%\hline
%$ Ztt $  & $ -7_\cdot 39\times 10^{-5} -  4_\cdot 39 \times 10^{-4} \ i $  \\
%\hline
%$ bbW $  & $ -1_\cdot 29\times 10^{-5} +  3_\cdot 79 \times 10^{-4} \ i $  \\
%\hline
%$ bWW $  & $ -3_\cdot 39\times 10^{-3} +  3_\cdot 08 \times 10^{-3} \ i $  \\
%\hline
%$ h_{0}tt $  & $ 3_\cdot 10\times 10^{-5} +  3_\cdot 97 \times 10^{-4} \ i $  \\
%\hline
%$ \textbf{Total} $  & $ -4_\cdot 62\times 10^{-3} + 5_\cdot 25\times 10^{-3}\ i $  \\
%\hline
%\end{tabular}
%%\caption{Continuation of Table I. Contributions of the BLHM to $a^{W}_{t}$.
%%\label{CN3}}
%\end{table}

\vspace{3.5cm}

\begin{center}
{\bf Acknowledgements}
\end{center}

E. C. A. appreciates the post-doctoral stay. A. G. R. thank SNI and PROFEXCE (M\'exico).

\vspace{3cm}

%\newpage

\appendix

%\begin{center}
\section{Feynman rules for the BLHM}
%\end{center}

In this appendix we present the complete set of Feynman rules for the BLHM involved in our calculation
for the AWMDM of the top-quark.

It is convenient to define the following useful notation:

\begin{eqnarray}
c_{\beta} &=& \cos \beta, \\
 s_{\beta} &=& \sin \beta, \\
s_{\alpha} &=& \sin \alpha, \\
c_{\alpha} &=& \cos \alpha.
\end{eqnarray}

\begin{eqnarray}
u_{i} &=& u,c.\\
d_{i} &=& d,s. \\
e_{i} &=& e,\mu, \tau. \\
\nu_{i} &=& \nu_{e}, \nu_{\mu}, \nu_{\tau}.
\end{eqnarray}

\begin{eqnarray}
y_{f }&=& \frac{m_f}{v \sin \beta} \left(1-\frac{v^{2}}{3 f^{2}} \right)^{-1/2}.
\end{eqnarray}

\begin{table}[H]
\caption{Feynman rules for the BLHM involving the pseudoscalar $A_0$.
\label{FeyRul-A0}}
\begin{tabular}{|c|p{14.3cm}|}
\hline
\hline
\textbf{Vertex} & \textbf{Feynman rules} \\
\hline
\hline
$A_{0}t\bar{t}$  & $-\frac{3 i c_{\beta} (P_{L}-P_{R}) y_{1} y_{2}
   y_{3}}{\sqrt{y_{1}^2+y_{2}^2}
   \sqrt{y_{1}^2+y_{3}^2}}+\frac{2 i c_{\beta}
   (P_{L}-P_{R}) s_{\beta} v y_{2}
   \left(y_{3}^2-2 y_{1}^2\right) y_{3}}{f
   \sqrt{y_{1}^2+y_{2}^2} \left(y_{1}^2+y_{3}^2\right)}   \hfill \break +   \frac{3 i c_{\beta} (P_{L}-P_{R}) s_{\beta}^2
   v^2 y_{1} y_{2} \left(4 y_{1}^6+4 \left(2
  y_{2}^2+y_{3}^2\right) y_{1}^4+\left(4 y_{2}^4+8
  y_{3}^2 y_{2}^2-2 \sqrt{y_{1}^2+y_{3}^2}\right)
  y_{1}^2+y_{2}^2 \left(4 y_{2}^2
   y_{3}^2+\sqrt{y_{1}^2+y_{3}^2}\right)\right)
   y_{3}^3}{2 f^2 \left(y_{1}^2+y_{2}^2\right)^{5/2}
   \left(y_{1}^2+y_{3}^2\right)^{5/2}} $ \\
\hline
 $A_{0} T\bar{T}$  & $  \frac{i c_{\beta} (P_{L}-P_{R}) s_{\beta} v
   y_{1}^2 \left(\left(y_{2}^2+2 y_{3}^2\right)
   y_{1}^4+\left(8 y_{2}^2 y_{3}^2-2 y_{3}^4\right)
   y_{1}^2-y_{2}^2 \left(y_{2}^4+3 y_{3}^2
   y_{2}^2-7 y_{3}^4\right)\right)}{f
   \left(y_{1}^2+y_{2}^2\right)^{3/2}
   \left(y_{2}^2-y_{3}^2\right)
   \left(y_{1}^2+y_{3}^2\right)} $ \\
\hline
$A_{0} T_{5}\bar{T}_{5}$  & $ \frac{9 i c_{\beta} (P_{L}-P_{R}) s_{\beta} v
   y_{1}^2 y_{2}^2 y_{3}^2}{f
   \left(y_{1}^2+y_{3}^2\right)^{3/2}
   \left(y_{3}^2-y_{2}^2\right)} $ \\
\hline
$A_{0} T_{6}\bar{T}_{6}$  & $ \frac{i c_{\beta} (P_{R}-P_{L}) s_{\beta} v
   y_{1}}{f} $ \\
\hline
$A_{0} T^{2/3}\bar{T}^{2/3}$  & $  \frac{i c_{\beta} (P_{L}-P_{R}) s_{\beta} v
   y_{1} \left(3 y_{1}^2-5 y_{3}^2\right)}{f
   \left(y_{1}^2+y_{3}^2\right)} $ \\
\hline
$A_{0} b\bar{b}$  & $  i c_{\beta} (P_{L}-P_{R}) y_{b} $ \\
\hline
\end{tabular}
\end{table}

\begin{table}[H]
\caption{Feynman rules for the BLHM involving the scalar $h_0$.
\label{FeyRul-h0}}
\begin{tabular}{|c|p{14.3cm}|}
\hline
\hline
\textbf{Vertex} & \textbf{Feynman rules} \\
\hline
\hline
$h_{0} T\bar{T}$  & $  \frac{c_{\alpha}  s_{\beta} v
   y_{1}^2 \left(\left(y_{2}^2+2 y_{3}^2\right)
   y_{1}^4+\left(8 y_{2}^2 y_{3}^2-2 y_{3}^4\right)
   y_{1}^2-y_{2}^2 \left(y_{2}^4+3 y_{3}^2
   y_{2}^2-7 y_{3}^4\right)\right)}{f
   \left(y_{1}^2+y_{2}^2\right)^{3/2}
   \left(y_{2}^2-y_{3}^2\right)
   \left(y_{1}^2+y_{3}^2\right)} $ \\
\hline
$h_{0} T_{5}\bar{T}_{5}$  & $ \frac{9 c_{\alpha}  s_{\beta} v
   y_{1}^2 y_{2}^2 y_{3}^2}{f
   \left(y_{1}^2+y_{3}^2\right)^{3/2}
   \left(y_{3}^2-y_{2}^2\right)} $ \\
\hline
$h_{0} T_{6}\bar{T}_{6}$  & $ -\frac{c_{\beta}  s_{\alpha} v
   y_{1}}{f} $ \\
\hline
$h_{0} T^{2/3}\bar{T}^{2/3}$  & $ -\frac{ v y_{1} \left(4 c_{\alpha}
   s_{\beta} \left(y_{1}^2-y_{3}^2\right)+\text{c$\beta
   $} s_{\alpha} \left(y_{1}^2+y_{3}^2\right)\right)}{f
   \left(y_{1}^2+y_{3}^2\right)} $ \\
\hline
$h_{0} t \bar{T}$  & $ \frac{c_{\alpha}
   P_{R} \left(y_{2}^2-2 y_{1}^2\right)
   y_{3}}{\sqrt{y_{1}^2+y_{2}^2}
   \sqrt{y_{1}^2+y_{3}^2}} + \frac{c_{\alpha}
   s_{\beta} y_{1} \left(2 P_{R} y_{3}
   \left(-y_{1}^4+y_{3}^2 y_{1}^2+y_{2}^4+y_{2}^2
   y_{3}^2\right)+P_{L} y_{2} \left(2 y_{1}^4+\left(2
   y_{2}^2+5 y_{3}^2\right) y_{1}^2-4 y_{2}^2
   y_{3}^2\right)\right) v}{f
   \left(y_{1}^2+y_{2}^2\right)^{3/2}
   \left(y_{1}^2+y_{3}^2\right)} $ \\
\hline
\end{tabular}
\end{table}

\begin{table}[H]
\caption{Feynman rules for the BLHM involving the scalar $H_0$.
\label{FeyRul-H0}}
\begin{tabular}{|c|p{14.3cm}|}
\hline
\hline
\textbf{Vertex} & \textbf{Feynman rules} \\
\hline
\hline
$H_{0} t\bar{t}$  & $  \frac{3
   s_{\alpha} y_{1} y_{2}
   y_{3}}{\sqrt{y_{1}^2+y_{2}^2}
   \sqrt{y_{1}^2+y_{3}^2}}  -\frac{2
    s_{\alpha} s_{\beta} y_{2}
   y_{3} \left(y_{3}^2-2 y_{1}^2\right) v}{f
   \sqrt{y_{1}^2+y_{2}^2}
   \left(y_{1}^2+y_{3}^2\right)}   - \frac{ y_{1} y_{2} y_{3} v^2}{2 f^2
   \left(y_{1}^2+y_{2}^2\right)^{5/2}
   \left(y_{1}^2+y_{3}^2\right)^{5/2}}
   \bigg(s_{\alpha}  \Big(14 y_{1}^8 +4 \left(7
   y_{2}^2+4 y_{3}^2\right) y_{1}^6 \hfill \break +2 \left(7
   y_{2}^4+16 y_{3}^2 y_{2}^2+y_{3}^4+2
   \sqrt{y_{1}^2+y_{3}^2}\right) y_{1}^4  -2 \big(-8
   y_{3}^2 y_{2}^4  +\left(\sqrt{y_{1}^2+y_{3}^2}-2
   y_{3}^4\right) y_{2}^2  \hfill \break  + y_{3}^2
   \sqrt{y_{1}^2+y_{3}^2}\big) y_{1}^2 +y_{2}^2
   y_{3}^2 \left(2 y_{2}^2 y_{3}^2+\sqrt{y_{1}^2+y_{3}^2}\right)\Big)
   s_{\beta}^2   -2 c_{\alpha} c_{\beta}
   \left(y_{1}^2+y_{3}^2\right) \hfill \break \times \left(6 y_{1}^6+6 \left(2
   y_{2}^2+y_{3}^2\right) y_{1}^4+2 \left(3 y_{2}^4+6
   y_{3}^2 y_{2}^2+\sqrt{y_{1}^2+y_{3}^2}\right)
   y_{1}^2+6 y_{2}^4 y_{3}^2-y_{2}^2
   \sqrt{y_{1}^2+y_{3}^2}\right) s_{\beta} \hfill \break   +2
   c_{\beta}^2 s_{\alpha}
   \left(y_{1}^2+y_{2}^2\right)^2
   \left(y_{1}^2+y_{3}^2\right)^2\bigg)   $ \\
\hline
$H_{0} T\bar{T}$  & $ -\frac{ s_{\alpha} s_{\beta} v
   y_{1}^2 \left(\left(y_{2}^2+2 y_{3}^2\right)
   y_{1}^4+\left(8 y_{2}^2 y_{3}^2-2 y_{3}^4\right)
   y_{1}^2-y_{2}^2 \left(y_{2}^4+3 y_{3}^2
   y_{2}^2-7 y_{3}^4\right)\right)}{f
   \left(y_{1}^2+y_{2}^2\right)^{3/2}
   \left(y_{2}^2-y_{3}^2\right)
   \left(y_{1}^2+y_{3}^2\right)} $ \\
\hline
$H_{0} T_{5}\bar{T}_{5}$  & $  -\frac{9  s_{\alpha} s_{\beta} v
   y_{1}^2 y_{2}^2 y_{3}^2}{f
   \left(y_{1}^2+y_{3}^2\right)^{3/2}
   \left(y_{3}^2-y_{2}^2\right)} $ \\
\hline
$H_{0} T_{6}\bar{T}_{6}$  & $ -\frac{c_{\alpha} c_{\beta}  v
   y_{1}}{f} $ \\
\hline
\end{tabular}
\end{table}

\begin{table}[H]
\caption{Continuation of Table~\ref{FeyRul-H0}.
\label{FeyRul-H00}}
\begin{tabular}{|c|p{14.3cm}|}
\hline
\hline
\textbf{Vertex} & \textbf{Feynman rules} \\
\hline
\hline
$H_{0} t\bar{T}_{6}$  & $ -\frac{2 c_{\alpha} P_{L} y_{1} y_{2}}{\sqrt{y_{1}^2+y_{2}^2}}  +\frac{P_{R}v y_{1}
   y_{3}}{2 f \left(y_{1}^2+y_{3}^2\right)^{3/2}} \hfill \break \times \left(\frac{c_{\alpha} s_{\beta} \left(4
   y_{1}^6+4 \left(2 y_{2}^2+y_{3}^2\right) y_{1}^4+2
   \left(2 y_{2}^4+4 y_{3}^2
   y_{2}^2+\sqrt{y_{1}^2+y_{3}^2}\right) y_{1}^2+4
   y_{2}^4 y_{3}^2-y_{2}^2
   \sqrt{y_{1}^2+y_{3}^2}\right)}{\left(y_{1}^2+y_{2}^
   2\right)^2}-8 c_{\beta} s_{\alpha}
   \left(y_{1}^2+y_{3}^2\right)\right)  \hfill \break - \frac{v^2}{3
   f^2 \sqrt{y_{1}^2+y_{2}^2}
   \left(y_{1}^2+y_{3}^2\right)} \bigg(2 c_{\beta} s_{\alpha} s_{\beta}
   \left(3 P_{R} \sqrt{y_{1}^2+y_{2}^2} y_{3}
   \left(y_{1}^2-3 y_{3}^2\right)+P_{L} \left(5 y_{1}
   y_{2} y_{3}^2-4 y_{1}^3
   y_{2}\right)\right)+c_{\alpha} \left(12 P_{L}
   y_{1} y_{2} \left(y_{1}^2+y_{3}^2\right)
   c_{\beta}^2+s_{\beta}^2 \left(P_{L} y_{1}
   y_{2} \left(4 y_{1}^2-5 y_{3}^2\right)+6 P_{R}
   \sqrt{y_{1}^2+y_{2}^2} y_{3}
   \left(y_{1}^2+y_{3}^2\right)\right)\right)\bigg)  $ \\
\hline
\end{tabular}
\end{table}

\begin{table}[H]
\caption{Feynman rules for the BLHM involving the scalar $\sigma$.
\label{FeyRul-sig}}
\begin{tabular}{|c|p{14.3cm}|}
\hline
\hline
\textbf{Vertex} & \textbf{Feynman rules} \\
\hline
\hline
$\sigma t\bar{t}$  & $ -\frac{3 c_{\beta}
    v y_{1} y_{2} y_{3}}{\sqrt{2} f
   \sqrt{y_{1}^2+y_{2}^2} \sqrt{y_{1}^2+y_{3}^2}} + \frac{\sqrt{2} c_{\beta} s_{\beta} v^2 y_{2} y_{3} \left(2 y_{1}^2+5
   y_{3}^2\right)}{f^2 \sqrt{y_{1}^2+y_{2}^2}
   \left(y_{1}^2+y_{3}^2\right)} $ \\
\hline
$\sigma T\bar{T}$  & $ -\frac{c_{\beta}  s_{\beta} v^2
   y_{1}^2 \left(3 \left(y_{2}^2+2 y_{3}^2\right)
   y_{1}^4+2 \left(2 y_{2}^4+2 y_{3}^2 y_{2}^2+5
   y_{3}^4\right) y_{1}^2+y_{2}^2 \left(y_{2}^4+7
   y_{3}^2 y_{2}^2+y_{3}^4\right)\right)}{\sqrt{2} f^2
   \left(y_{1}^2+y_{2}^2\right)^{3/2}
   \left(y_{2}^2-y_{3}^2\right)
   \left(y_{1}^2+y_{3}^2\right)} $ \\
\hline
$\sigma T_{5}\bar{T}_{5}$  & $  \frac{c_{\beta}  s_{\beta} v^2
   y_{3}^2 \left(-6 y_{1}^4+\left(7 y_{2}^2-4
   y_{3}^2\right) y_{1}^2+2 y_{3}^4-2 y_{2}^2
   y_{3}^2\right)}{\sqrt{2} f^2
   \left(y_{1}^2+y_{3}^2\right)^{3/2}
   \left(y_{3}^2-y_{2}^2\right)} $ \\
\hline
$\sigma T_{6}\bar{T}_{6}$  & $  -\frac{3 c_{\beta}  s_{\beta} v^2
   y_{1}}{\sqrt{2} f^2} $ \\
\hline
$\sigma T^{2/3}\bar{T}^{2/3}$  & $ -\frac{c_{\beta}  s_{\beta} v^2
   y_{1} \left(y_{1}^2-7 y_{3}^2\right)}{\sqrt{2} f^2
   \left(y_{1}^2+y_{3}^2\right)} $ \\
\hline
$\sigma b\bar{b}$  & $  -\frac{c_{\beta}  v \text{yb}}{\sqrt{2}
   f} $ \\
\hline
$\sigma t\bar{T}$  & $  \frac{c_{\beta}
   P_{R} \left(y_{2}^2-2 y_{1}^2\right) y_{3}
   v}{\sqrt{2} f \sqrt{y_{1}^2+y_{2}^2}
   \sqrt{y_{1}^2+y_{3}^2}} + \frac{c_{\beta} s_{\beta} y_{1} \left(2 P_{R}
   \left(y_{1}^2+y_{2}^2\right) y_{3} \left(3
   y_{1}^2+y_{2}^2+5 y_{3}^2\right)+P_{L} \left(-2
   y_{2} y_{1}^4+\left(13 y_{2} y_{3}^2-2
   y_{2}^3\right) y_{1}^2+4 y_{2}^3
   y_{3}^2\right)\right) v^2}{\sqrt{2} f^2
   \left(y_{1}^2+y_{2}^2\right)^{3/2}
   \left(y_{1}^2+y_{3}^2\right)}  $ \\
\hline
\end{tabular}
\end{table}

\begin{table}[H]
\caption{Feynman rules for the BLHM involving the scalar $\phi^{0}$.
\label{FeyRul-p0}}
\begin{tabular}{|c|p{14.3cm}|}
\hline
\hline
\textbf{Vertex} & \textbf{Feynman rules} \\
\hline
\hline
$\phi^{0} t\bar{t}$  & $ \frac{3 i (P_{L}-P_{R}) s_{\beta} v y_{1}
   y_{2} y_{3}}{2 f \sqrt{y_{1}^2+y_{2}^2}
   \sqrt{y_{1}^2+y_{3}^2}}-\frac{i (P_{L}-P_{R})
   s_{\beta}^2 v^2 y_{2} y_{3} \left(y_{3}^2-2
   y_{1}^2\right)}{f^2 \sqrt{y_{1}^2+y_{2}^2}
   \left(y_{1}^2+y_{3}^2\right)} $ \\
\hline
$\phi^{0} T\bar{T}$  & $ \frac{i (P_{L}-P_{R}) v^2}{2 f^2 \left(y_{1}^2+y_{2}^2\right)^{3/2}
   \left(y_{2}^2-y_{3}^2\right)
   \left(y_{1}^2+y_{3}^2\right) } \Big(\left(3 y_{1}^4+2
   y_{2}^2 y_{1}^2-y_{2}^4\right)
   \left(y_{2}^2-y_{3}^2\right)
   \left(y_{1}^2+y_{3}^2\right) c_{\beta
   }^2 \hfill \break +s_{\beta}^2 \left(\left(2
   y_{2}^2+y_{3}^2\right) y_{1}^6  +\left(2
   y_{2}^4-y_{3}^2 y_{2}^2+5 y_{3}^4\right)
   y_{1}^4+\left(6 y_{2}^4 y_{3}^2-3 y_{2}^2
   y_{3}^4\right) y_{1}^2+y_{2}^4 y_{3}^4-y_{2}^6
   y_{3}^2\right)\Big) $ \\
\hline
$\phi^{0} T_{5}\bar{T}_{5}$  & $ -\frac{9 i (P_{L}-P_{R}) s_{\beta}^2 v^2 y_{1}^2
   y_{2}^2 y_{3}^2}{2 f^2
   \left(y_{1}^2+y_{3}^2\right)^{3/2}
   \left(y_{3}^2-y_{2}^2\right)} $ \\
\hline
$\phi^{0} T_{6}\bar{T}_{6}$  & $ -\frac{3 i c_{\beta}^2 (P_{L}-P_{R}) v^2 y_{1}}{2
   f^2} $ \\
\hline
$\phi^{0} T^{2/3}\bar{T}^{2/3}$  & $  -\frac{i (P_{L}-P_{R}) s_{\beta}^2 v^2 y_{1}
   \left(y_{1}^2-7 y_{3}^2\right)}{2 f^2
   \left(y_{1}^2+y_{3}^2\right)} $ \\
\hline
$\phi^{0} b\bar{b}$  & $ -\frac{i (P_{L}-P_{R}) s_{\beta} v \text{yb}}{2 f} $ \\
\hline
\end{tabular}
\end{table}

\begin{table}[H]
\caption{Feynman rules for the BLHM involving the scalars $\eta^{0}$, $H^+$, $\phi^+$ and $\eta^+$.
%Set of Yukawa couplings in the BLHM for the scalars $\eta^{0}$, $H^+$, $\phi^+$ and $\eta^+$.
\label{FeyRul-eta0}}
\begin{tabular}{|c|p{14.3cm}|}
\hline
\hline
\textbf{Vertex} & \textbf{Feynman rules} \\
\hline
\hline
$\eta^{0} t\bar{t}$  & $ -\frac{3 i
   (P_{L}-P_{R}) s_{\beta} v y_{1} y_{2}
   y_{3}}{2 f \sqrt{y_{1}^2+y_{2}^2}
   \sqrt{y_{1}^2+y_{3}^2}} + \frac{i (P_{L}-P_{R}) s_{\beta}^2 v^2 y_{2}
   y_{3} \left(y_{3}^2-2 y_{1}^2\right)}{f^2
   \sqrt{y_{1}^2+y_{2}^2}
   \left(y_{1}^2+y_{3}^2\right)} $ \\
\hline
$\eta^{0} T\bar{T}$  & $ -\frac{i (P_{L}-P_{R}) v^2 }{2 f^2
   \left(y_{1}^2+y_{2}^2\right)^{3/2}
   \left(y_{2}^2-y_{3}^2\right)
   \left(y_{1}^2+y_{3}^2\right)} \Big(\left(3 y_{1}^4+2
   y_{2}^2 y_{1}^2-y_{2}^4\right)
   \left(y_{2}^2-y_{3}^2\right)
   \left(y_{1}^2+y_{3}^2\right) c_{\beta}^2 \hfill \break +s_{\beta}^2 \left(\left(2
   y_{2}^2+y_{3}^2\right) y_{1}^6+\left(2
   y_{2}^4-y_{3}^2 y_{2}^2+5 y_{3}^4\right)
   y_{1}^4+\left(6 y_{2}^4 y_{3}^2-3 y_{2}^2
   y_{3}^4\right) y_{1}^2+y_{2}^4 y_{3}^4-y_{2}^6
   y_{3}^2\right)\Big) $ \\
\hline
$\eta^{0} T_{5}\bar{T}_{5}$  & $  \frac{9 i (P_{L}-P_{R}) s_{\beta}^2 v^2 y_{1}^2
   y_{2}^2 y_{3}^2}{2 f^2
   \left(y_{1}^2+y_{3}^2\right)^{3/2}
   \left(y_{3}^2-y_{2}^2\right)} $ \\
\hline
$\eta^{0} T_{6}\bar{T}_{6}$  & $ \frac{3 i c_{\beta}^2 (P_{L}-P_{R}) v^2 y_{1}}{2
   f^2} $ \\
\hline
$\eta^{0} T^{2/3}\bar{T}^{2/3}$  & $ \frac{i (P_{L}-P_{R}) s_{\beta}^2 v^2 y_{1}
   \left(y_{1}^2-7 y_{3}^2\right)}{2 f^2
   \left(y_{1}^2+y_{3}^2\right)} $ \\
\hline
$\eta^{0} b\bar{b}$  & $  \frac{i (P_{L}-P_{R}) s_{\beta} v \text{yb}}{2 f} $ \\
\hline
\end{tabular}
\end{table}

%%%%%%%%%%%%%%%%%%%%%%%%%%%%%%%%%%%%%%%
%%%%%%%%%%%%%%%%%%%%%%%%%%%%%%%%%%%%%%%%%%%%%%

\begin{table}[H]
\caption{Continuation of Table~\ref{FeyRul-eta0}.
%Set of Yukawa couplings in the BLHM for the scalars $\eta^{0}$, $H^+$, $\phi^+$ and $\eta^+$.
\label{FeyRul-eta0-0}}
\begin{tabular}{|c|p{14.3cm}|}
\hline
\hline
\textbf{Vertex} & \textbf{Feynman rules} \\
\hline
\hline
$H^{-} \bar{B} t$  & $ \frac{\sqrt{2} \text{c$\beta
   $} P_{R} \left(y_{2}^2-2 y_{1}^2\right)
   y_{3}}{\sqrt{y_{1}^2+y_{2}^2}
   \sqrt{y_{1}^2+y_{3}^2}} + \frac{2 \sqrt{2} c_{\beta} P_{R} s_{\beta} v
   y_{1} \left(-y_{1}^2+y_{2}^2+y_{3}^2\right)
   y_{3}}{f \sqrt{y_{1}^2+y_{2}^2}
   \left(y_{1}^2+y_{3}^2\right)}  $ \\
\hline
$H^{-} \bar{B} T$  & $  \frac{\sqrt{2} c_{\beta} P_{R} s_{\beta} v
   y_{1}^2 \left(y_{1}^4+5 y_{3}^2 y_{1}^2-y_{2}^2
   \left(y_{2}^2+4 y_{3}^2\right)\right)}{f
   \left(y_{1}^2+y_{2}^2\right)^{3/2}
   \left(y_{1}^2+y_{3}^2\right)} $ \\
\hline
$H^{-} \bar{B} T_{5}$  & $\frac{\sqrt{2} c_{\beta} P_{R} y_{1}
   \left(-y_{1}^2+y_{2}^2+y_{3}^2\right)}{\sqrt{y_{1}^2
   +y_{2}^2} \sqrt{y_{1}^2+y_{3}^2}}  $ \\
\hline
$H^{-} \bar{B} T_{6}$  & $ \frac{\sqrt{2} P_{R} s_{\beta}
   \left(y_{1}^2-y_{2}^2\right)}{\sqrt{y_{1}^2+y_{2}^2}
   } $ \\
\hline
$H^{-} \bar{B} T^{2/3}$  & $ -\frac{2 \sqrt{2} c_{\beta} P_{R} s_{\beta} v
   \left(y_{1}^4-\left(y_{2}^2+3 y_{3}^2\right)
   y_{1}^2+y_{2}^2 y_{3}^2\right)}{f
   \sqrt{y_{1}^2+y_{2}^2} \left(y_{1}^2+y_{3}^2\right)}  $ \\
\hline
\hline
$\phi^{+} \bar{t} b$  & $  \frac{i s_{\beta} v}{\sqrt{2} f} \left(\frac{3 P_{L} y_{1}
   y_{2} y_{3}}{\sqrt{y_{1}^2+y_{2}^2}
   \sqrt{y_{1}^2+y_{3}^2}}-P_{R}
   \text{yb}\right)-\frac{i \sqrt{2} P_{L}
   s_{\beta}^2 v^2 y_{2} y_{3} \left(y_{3}^2-2
   y_{1}^2\right)}{f^2 \sqrt{y_{1}^2+y_{2}^2}
   \left(y_{1}^2+y_{3}^2\right)} $ \\
\hline
\end{tabular}
\end{table}

\begin{table}[H]
\caption{Feynman rules for the BLHM involving the current sector.
\label{FeyRul-current-1}}
\begin{tabular}{|c|p{14.3cm}|}
\hline
\textbf{Vertex} & \textbf{Feynman rules} \\
\hline
\hline
$\gamma \bar{T} T$  & $\frac{2}{3} i g s_{W} \gamma ^{\mu
   }  +  \frac{2 i g s_{W}  v^2}{3
   f^2} \bigg( c_{\beta}^2  \gamma ^{\mu }.P_{R}   +\frac{y_{1}^2 \left(4 c_{\beta}^2
   \left(y_{2}^2-y_{3}^2\right)^2+s_{\beta}^2 \left(2
   y_{2}^2+y_{3}^2\right)^2\right) \gamma ^{\mu
   }.P_{L}}{\left(y_{1}^2+y_{2}^2\right)
   \left(y_{2}^2-y_{3}^2\right)^2}+\frac{s_{\beta}^2
   \left(y_{2}^2-2 y_{1}^2\right)^2 y_{3}^2 \gamma ^{\mu
   }.P_{R}}{\left(y_{1}^2+y_{2}^2\right)^2
   \left(y_{1}^2+y_{3}^2\right)}\bigg) $ \\
\hline
$\gamma \bar{T_{5}} T_{5}$  & $ \frac{2}{3} i g s_{W}
  \gamma ^{\mu }+  \frac{3 i g s_{W} s_{\beta}^2 v^2 y_{1}^2
   \left(\left(6 y_{3}^4-2 y_{2}^2 y_{3}^2+y_{1}^2
   \left(y_{2}^2+2 y_{3}^2\right)\right) \gamma ^{\mu
   }.P_{L}-\left(y_{2}^2-4 y_{3}^2\right)
   \left(y_{1}^2+y_{3}^2\right) \gamma ^{\mu
   }.P_{R}\right) y_{2}^2}{2 f^2
   \left(y_{2}^2-y_{3}^2\right)^2
   \left(y_{1}^2+y_{3}^2\right)^2} $ \\
\hline
$\gamma \bar{T_{6}} T_{6}$  & $ \frac{2}{3}
   i g s_{W} \gamma ^{\mu } +  \frac{i c_{\beta}^2 g s_{W} \gamma ^{\mu } \left(4 P_{L}+P_{R}\right) v^2}{2 f^2} $ \\
\hline
$\gamma \bar{T }^{2/3} T^{2/3}$  & $\frac{2}{3} i g s_{W} \gamma ^{\mu
   } +  \frac{2 i g s_{W} s_{\beta}^2   \gamma ^{\mu }\left(
P_{L}+4 P_{R}\right) v^2}{3
   f^2} $ \\
\hline
$\gamma \bar{T }^{5/3} T^{5/3}$  & $ \frac{5}{3} i g s_{W} \gamma ^{\mu
   } $ \\
\hline
$\gamma \bar{B } B$  & $  -\frac{1}{3} i g s_{W} \gamma ^{\mu
   }  $ \\
\hline
\end{tabular}
\end{table}

\begin{table}[H]
\caption{Continuation of Table~\ref{FeyRul-current-1}.
\label{FeyRul-current-2}}
\begin{tabular}{|c|p{14.3cm}|}
\hline
\textbf{Vertex} & \textbf{Feynman rules} \\
\hline
\hline
$Z \bar{t} T_{6}$  & $ -\frac{i c_{\beta} g
   \left(3 c_{W}^2+4 s_{W}^2\right) y_{2} v \gamma ^{\mu
   }.P_{L} }{3 c_{W} f \sqrt{y_{1}^2+y_{2}^2}}  -\frac{i c_{W} c_{\beta} g s_{\beta} \left(2
   y_{1}^2-y_{2}^2\right) y_{3}v^2 \gamma ^{\mu }.P_{R}
   }{4 f^2 \left(y_{1}^2+y_{2}^2\right)^2
   \left(y_{1}^2+y_{3}^2\right)} $ \\
\hline
$Z \bar{t } T^{2/3}$  & $ -\frac{i g \left(3
   c_{W}^2-4 s_{W}^2\right) s_{\beta} y_{3} v \gamma
   ^{\mu }.P_{R} }{3 c_{W} f \sqrt{y_{1}^2+y_{3}^2}} -\frac{2 i g s_{W}^2 s_{\beta}^2 v^2 \left(y_{2}
   \left(y_{3}^2-2 y_{1}^2\right) \gamma ^{\mu }.P_{L}+4
   y_{1} \sqrt{y_{1}^2+y_{2}^2} y_{3} \gamma ^{\mu
   }.P_{R}\right) }{3 c_{W} f^2
   \sqrt{y_{1}^2+y_{2}^2}
   \left(y_{1}^2+y_{3}^2\right)} $ \\
\hline
$Z \bar{b }B $  & $  \frac{i g s_{W}^2 y_{1} y_{2} \gamma ^{\mu }.PL}{6
   c_{W} \left(y_{1}^2+y_{2}^2\right)} $ \\
\hline
\hline
$Z' \bar{t} t$  & $  \frac{1}{2} i g \gamma ^{\mu }.P_{L}-\frac{i g s_{\beta}^2
   v^2 y_{3}^2 }{8 f^2 \left(y_{1}^2+y_{3}^2\right)^2} \left(\frac{\left(y_{2}^2-2
   y_{1}^2\right)^2}{\left(y_{1}^2+y_{2}^2\right)^4}-16
   \left(y_{1}^2+y_{3}^2\right)\right) \gamma ^{\mu
   }.P_{R} $ \\
\hline
$Z' \bar{T} T$  & $ \frac{1}{2} i g \gamma ^{\mu }.\left(P_{L}-P_{R}\right) $ \\
\hline
$Z' \bar{T_{5}} T_{5}$  & $  \frac{3 i g s_{\beta}^2 v^2 y_{1}^2 y_{2}^2
   \left(\left(5 y_{3}^4-2 y_{2}^2 y_{3}^2+y_{1}^2
   \left(y_{2}^2+2 y_{3}^2\right)\right) \gamma ^{\mu
   }.P_{L}+\left(y_{2}^2-4 y_{3}^2\right)
   \left(y_{1}^2+y_{3}^2\right) \gamma ^{\mu
   }.P_{R}\right)}{2 f^2 \left(y_{2}^2-y_{3}^2\right)^2
   \left(y_{1}^2+y_{3}^2\right)^2} $ \\
\hline
$Z' \bar{T_{6}} T_{6}$  & $ \frac{i c_{\beta}^2 g v^2 \left(4 \gamma ^{\mu
   }.P_{L}-\gamma ^{\mu }.P_{R}\right)}{2 f^2} $ \\
\hline
\end{tabular}
\end{table}

\begin{table}[H]
\caption{Continuation of Table~\ref{FeyRul-current-2}.
\label{FeyRul-current-3}}
\begin{tabular}{|c|p{14.3cm}|}
\hline
\textbf{Vertex} & \textbf{Feynman rules} \\
\hline
$W^{-} \bar{T }^{2/3} T^{5/3} $  & $  \frac{ig \gamma^{\mu}.P_{L} }{\sqrt{2}}  $ \\
\hline
$W^{-} \bar{ T}_{5} T^{5/3}$  & $  \frac{ig v s_{\beta} \gamma^{\mu}.P_{L} }{\sqrt{2} f}  $ \\
\hline
$W^{-} \bar{ B} T$  & $ \frac{ig \gamma^{\mu}.P_{L} }{\sqrt{2}}  $ \\
\hline
$W^{-} \bar{B} T_{5}$  & $  -\frac{i g s_{\beta} v y_{1} \left(2
   y_{2}^2+y_{3}^2\right) \gamma ^{\mu }.P_{L}}{\sqrt{2} f
   \sqrt{y_{1}^2+y_{2}^2} \left(y_{2}^2-y_{3}^2\right)}  $ \\
\hline
$W^{-} \bar{B} T_{6}$  & $  -\frac{i \sqrt{2} c_{\beta} g v y_{1} \gamma ^{\mu
   }.P_{L}}{f \sqrt{y_{1}^2+y_{2}^2}} $ \\
\hline
$W^{-} \bar{ b} T_{5} $  & $  -\frac{i g s_{\beta} v y_{2} \left(2
   y_{1}^2-y_{3}^2\right) \gamma ^{\mu }.P_{L}}{\sqrt{2} f
   \sqrt{y_{1}^2+y_{2}^2} \left(y_{1}^2+y_{3}^2\right)} $ \\
\hline
$W^{-} \bar{ b} T_{6}$  & $ -\frac{i \sqrt{2} c_{\beta} g v y_{2} \gamma ^{\mu
   }.P_{L}}{f \sqrt{y_{1}^2+y_{2}^2}} $ \\
\hline
\end{tabular}
\end{table}

\begin{table}[H]
\caption{Continuation of Table~\ref{FeyRul-current-3}.
\label{FeyRul-current-04}}
\begin{tabular}{|c|p{14.3cm}|}
\hline
\textbf{Vertex} & \textbf{Feynman rules} \\
\hline
\hline
$Z' \bar{u_i} u_i$  & $ \frac{1}{2} i g \gamma^{\mu} P_{L}  $ \\
\hline
$Z' \bar{d_i} d_i$  & $ -\frac{1}{2} i g \gamma^{\mu} P_{L}  $ \\
\hline
$Z' \bar{e_i} e_i$  & $ \frac{1}{2} i g \gamma^{\mu} P_{L}   $ \\
\hline
$Z' \bar{\nu_i} \nu_i$  & $ -\frac{1}{2} i g \gamma^{\mu} P_{L}   $ \\
\hline
$W'^{+} \bar{u_i} d_i$  & $ \frac{ig}{\sqrt{2}}  \gamma^{\mu} P_{L}  $ \\
\hline
$W'^{+} \bar{\nu_i} e_i$  & $ - \frac{ig}{\sqrt{2}}  \gamma^{\mu} P_{L}   $ \\
\hline
\end{tabular}
\end{table}

\begin{table}[H]
\caption{Feynman rules for BLHM involving the Yukawa sector (light fermions).
\label{FeyRul-current-014}}
\begin{tabular}{|c|p{14.3cm}|}
\hline
\textbf{Vertex} & \textbf{Feynman rules} \\
\hline
\hline
$ h_0 \bar{u}_i u_i $  & $ c_{\alpha} y_{u_i}-\frac{\left(8 c_{\beta}
   s_{\alpha} s_{\beta} v^2+4 c_{\alpha}
   \left(c_{\beta}^2 v^2+3 s_{\beta}^2 v^2\right)\right)
   y_{u_i}}{12 f^2}   $  \\
\hline
$ h_0 \bar{d}_i d_i $  & $   c_{\alpha} y_{d_i}-\frac{\left(8 c_{\beta}
   s_{\alpha} s_{\beta} v^2+4 c_{\alpha}
   \left(c_{\beta}^2 v^2+3 s_{\beta}^2 v^2\right)\right)
   y_{d_i}}{12 f^2} $  \\
\hline
$ H_0 \bar{u}_i u_i $  & $-s_{\alpha}y_{u_i} +  \frac{\left(4 s_{\alpha} \left(c_{\beta}^2 v^2+3
   s_{\beta}^2 v^2\right)-8 c_{\alpha} c_{\beta}
   s_{\beta} v^2\right) y_{u_i}}{12 f^2}  $  \\
\hline
$ H_0 \bar{d}_i d_i $  & $ -s_{\alpha}y_{d_i}  + \frac{\left(4 s_{\alpha} \left(c_{\beta}^2 v^2+3
   s_{\beta}^2 v^2\right)-8 c_{\alpha} c_{\beta}
   s_{\beta} v^2\right) y_{d_i}}{12 f^2} $  \\
\hline
$ A_0 \bar{u}_i u_i $  & $  -i c_{\beta} y_{u_i} $  \\
\hline
$ A_0 \bar{d}_i d_i $  & $  -i c_{\beta} y_{d_i} $  \\
\hline
$ \sigma \bar{u}_i u_i $  & $  \frac{c_{\beta} v y_{u_i}}{\sqrt{2} f} $  \\
\hline
$ \sigma \bar{d}_i d_i $  & $  \frac{c_{\beta} v y_{d_i}}{\sqrt{2} f} $  \\
\hline
\hline
$ \eta^{0} \bar{u}_i u_i $  & $  -\frac{i s_{\beta} v y_{u_i}}{2 f} $  \\
\hline
$ \eta^{0} \bar{d}_i d_i $  & $  -\frac{i s_{\beta} v y_{d_i}}{2 f} $  \\
\hline
$ \phi^{0} \bar{u}_i u_i $  & $  \frac{i s_{\beta} v y_{u_i}}{2 f} $  \\
\hline
$ \phi^{0} \bar{d}_i d_i $  & $ \frac{i s_{\beta} v y_{d_i}}{2 f}  $  \\
\hline
$ \eta^{+} \bar{u}_i d_i $  & $ \frac{i s_{\beta} v y_{d_i}}{\sqrt{2} f}P_{R} $  \\
\hline
$ \phi^{+} \bar{u}_i d_i $  & $  -\frac{i s_{\beta} v y_{d_i}}{\sqrt{2} f}P_{R} $  \\
\hline
$ H^{+} \bar{u}_i d_i $  & $   \sqrt{2} c_{\beta} y_{d_i} P_{R} $ \\
\hline
\end{tabular}
\end{table}

\begin{table}[H]
\caption{Continuation of Table~\ref{FeyRul-current-014}.
\label{FeyRul-current-4}}
\begin{tabular}{|c|p{14.3cm}|}
\hline
\textbf{Vertex} & \textbf{Feynman rules} \\
\hline
\hline
$ \sigma \bar{e}_i e_i $  & $   \frac{c_{\beta}  v y_{e_i}}{\sqrt{2} f}  $  \\
\hline
$ h_0 \bar{e}_i e_i $  & $  c_{\alpha} y_{e_i}-\frac{\left(16 c_{\beta}
   s_{\alpha} s_{\beta} v^2+c_{\alpha} \left(8
   c_{\beta}^2 v^2+24 s_{\beta}^2 v^2\right)\right)
   y_{e_i}}{24 f^2}   $  \\
\hline
$ H_0 \bar{e}_i e_i $  & $ -s_{\alpha} y_{e_i} +  \frac{\left(s_{\alpha} c_{\beta}^2-2 c_{\alpha}
   s_{\beta} c_{\beta}+3 s_{\alpha} s_{\beta}^2\right) v^2 y_{e_i}}{3 f^2}  $  \\
\hline
$ A_0 \bar{e}_i e_i $  & $   -i c_{\beta} y_{e_i}  $  \\
\hline
$ \eta^{0} \bar{e}_i e_i $  & $   -\frac{i s_{\beta} v y_{e_i}}{2 f}  $  \\
\hline
$ \phi^{0} \bar{e}_i e_i $  & $   \frac{i s_{\beta} v y_{e_i}}{2 f}  $  \\
\hline
$ H^{+} \bar{\nu}_i e_i $  & $   \sqrt{2} c_{\beta} y_{e_i}  P_{R} $  \\
\hline
$ \eta^{+} \bar{\nu}_i e_i $  & $  \frac{i s_{\beta} v y_{e_i}}{\sqrt{2} f}P_{R}   $  \\
\hline
$ \phi^{+} \bar{\nu}_i e_i $  & $  -\frac{i s_{\beta} v y_{e_i}}{\sqrt{2} f}P_{R}  $  \\
\hline
\hline
\end{tabular}
\end{table}

\vspace{5cm}

\section{  }

As additional information, in this appendix we give a summary of the numerical contributions of  particles that induce the AWMDM of the top-quark.

\begin{table}[H]
\caption{Expected sensitivity limits on the $a^W_t$ in the context of the BLHM with $\sqrt{q^{2}}=500\ \text{GeV}$,
\ $\text{m}_{A_0}=1 000\ \text{GeV}$, $\text{m}_{\eta_0}=100\ \text{GeV}$, $f=1 000\ \text{GeV}$, $F=4000\ \text{GeV}$
are represented. Here, any couplings are calculated while fixing the other couplings to zero. {\bf abc} denotes the different particles running in the loop of the vertex $Ztt$.
\label{CN1}}
\centering
\begin{tabular}{|c|c|}
\hline
\hline
\multicolumn{2}{|c|}{$\sqrt{q^{2}}=500\ \text{GeV}$,\ $\text{m}_{A_0}=1 000\ \text{GeV}$, $\text{m}_{\eta_0}=100\ \text{GeV}$, $f=1 000\ \text{GeV}$, $F=4000\ \text{GeV}$}\\
\hline
%\hline
 $\textbf{Couplings abc}$  & $\left( a^{W}_{t} \right)^{ \textbf{abc} } $  \\
\hline
\hline
$ Z'tt $  & $- 9_\cdot 39\times 10^{-7} - 1_\cdot 38\times 10^{-8}\ i$  \\
\hline
$ \sigma tt $  & $ 2_\cdot 52\times 10^{-8} + 1_\cdot 10\times 10^{-8}\ i$  \\
\hline
$H_{0} tt $  & $ 3_\cdot 69 \times 10^{-6} + 2_\cdot 65\times 10^{-6}\ i$  \\
\hline
$ A_{0} tt$  & $ -4_\cdot 29 \times 10^{-6} - 3_\cdot 11 \times 10^{-6}\ i$  \\
\hline
$ \phi^{0} tt $  & $ -1_\cdot 10 \times 10^{-6} - 1_\cdot 18\times 10^{-6}\ i$  \\
\hline
$ \eta^{0} tt $  & $ 7_\cdot 86 \times 10^{-8} - 7_\cdot 28\times 10^{-6}\ i$  \\
\hline
$ W' tt $  & $ 2_\cdot 33\times 10^{-6} + 6_\cdot 42\times 10^{-8}\ i$  \\
\hline
$H^{\pm} bb $  & $ -2 _\cdot 36 \times 10^{-6} - 6_\cdot 62\times 10^{-7}\ i$  \\
\hline
$ \phi^{\pm} bb $  & $ -8_\cdot 46\times 10^{-7} - 6_\cdot 07 \times 10^{-7}\ i$  \\
\hline
$ \eta^{\pm} bb $  & $ -1_\cdot 12 \times 10^{-6} - 2_\cdot 48\times 10^{-6}\ i$  \\
\hline
$ Z TT $  & $ -1_\cdot 55 \times 10^{-5}+0\ i $  \\
\hline
$ \gamma TT $  & $ - 6_\cdot 24 \times 10^{-7} +0\ i $  \\
\hline
$Z' TT $  & $ -4 _\cdot 83 \times 10^{-8}+0\ i $  \\
\hline
$ \sigma TT $  & $ 2_\cdot 11\times 10^{-9}+0\ i $  \\
\hline
$ h_{0} TT $  & $ -3_\cdot 43 \times 10^{-5}+0\ i $  \\
\hline
$ H_{0} TT $  & $- 1_\cdot 81 \times 10^{-6} + 0 i$  \\
\hline
$ A_{0} TT $  & $ 2_\cdot 54\times 10^{-6} + 0 i$  \\
\hline
$\phi^{0} TT $  & $ 3_\cdot 56 \times 10^{-7} + 0\ i$  \\
\hline
$ \eta^{0} TT $  & $ 6_\cdot 17 \times 10^{-7} +0\ i$  \\
\hline
$ Z T_{5} T_{5} $  & $ 5_\cdot 46 \times 10^{-6} +0\ i$  \\
\hline
$ \gamma T_{5} T_{5} $  & $ 4_\cdot 50 \times 10^{-7} +0\ i$  \\
\hline
\end{tabular}
%\caption{Expected sensitivity limits on the $a^W_t$ in the context of the BLHM with $\sqrt{q^{2}}=500\ \text{GeV}$,
%\ $\text{m}_{A_0}=1 000\ \text{GeV}$, $\text{m}_{\eta_0}=500\ \text{GeV}$, $f=1 000\ \text{GeV}$, $F=4000\ \text{GeV}$
%are represented. Here, any couplings are calculated while fixing the other couplings to zero.
%\label{CN1}}
\end{table}

\begin{table}[H]
\caption{Continuation of Table~\ref{CN1}.
\label{CN2}}
\centering
\begin{tabular}{|c|c|}
\hline
\hline
\multicolumn{2}{|c|}{$\sqrt{q^{2}}=500\ \text{GeV}$,\ $\text{m}_{A_0}=1 000\ \text{GeV}$, $\text{m}_{\eta_0}=100\ \text{GeV}$, $f=1 000\ \text{GeV}$, $F=4000\ \text{GeV}$}\\
\hline
\hline
$Z' T_{5} T_{5} $  & $ 6_\cdot 68\times 10^{-12} + 0\ i$  \\
\hline
$\sigma T_{5} T_{5} $  & $ -6_\cdot 22 \times 10^{-9}+0\ i$  \\
\hline
$ h_{0}T_{5} T_{5} $  & $ -2_\cdot 08\times 10^{-7} +0\ i$  \\
\hline
$ H_{0} T_{5} T_{5} $  & $ -1_\cdot 33 \times 10^{-8} +0\ i$  \\
\hline
$ A_{0} T_{5} T_{5} $  & $ 1_\cdot 23\times 10^{-8} +0\ i $  \\
\hline
$\phi^{0} T_{5} T_{5} $  & $ - 1_\cdot 11\times 10^{-11}+0\ i  $  \\
\hline
$\eta^{0} T_{5} T_{5} $  & $ -1 _\cdot 50 \times 10^{-11}+0\ i $  \\
\hline
$ Z T_{6} T_{6} $  & $ 2_\cdot 11 \times 10^{-6} +0\ i $  \\
\hline
$ \gamma T_{6} T_{6} $  & $ 1_\cdot 57 \times 10^{-9}+0\ i $  \\
\hline
$ Z' T_{6}T_{6} $  & $ 2_\cdot 43 \times 10^{-9} + 0 i$  \\
\hline
$ \sigma T_{6}T_{6} $  & $ -2_\cdot 06 \times 10^{-6} + 0 i$  \\
\hline
$h_{0} T_{6}T_{6} $  & $ 8_\cdot 87 \times 10^{-6} + 0\ i$  \\
\hline
$ H_{0} T_{6}T_{6} $  & $ 1_\cdot 73 \times 10^{-5} +0\ i$  \\
\hline
$ A_{0} T_{6}T_{6} $  & $ 6_\cdot 99 \times 10^{-6} +0\ i$  \\
\hline
$ \phi^{0} T_{6}T_{6} $  & $ -2_\cdot 99 \times 10^{-9} +0\ i$  \\
\hline
$\eta^{0} T_{6}T_{6} $  & $ -5_\cdot 51 \times 10^{-9} + 0\ i$  \\
\hline
$Z T^{2/3} T^{2/3}$  & $ 3 _\cdot 91 \times 10^{-5}+0\ i$  \\
\hline
$ \gamma T^{2/3} T^{2/3} $  & $ 7_\cdot 78 \times 10^{-6} +0\ i$  \\
\hline
$ Z' T^{2/3} T^{2/3} $  & $ 2_\cdot 45 \times 10^{-7} +0\ i$  \\
\hline
$ \sigma T^{2/3} T^{2/3} $  & $ -7_\cdot 81 \times 10^{-9} +0\ i $  \\
\hline
$h_{0} T^{2/3} T^{2/3} $  & $ 1_\cdot 76 \times 10^{-4}+0\ i  $  \\
\hline
$H_{0} T^{2/3} T^{2/3} $  & $ 3 _\cdot 27 \times 10^{-6}+0\ i $  \\
\hline
$ A_{0}T^{2/3} T^{2/3} $  & $ 1_\cdot 47 \times 10^{-5} +0\ i $  \\
\hline
$ \phi^{0} T^{2/3} T^{2/3} $  & $ -1_\cdot 52 \times 10^{-8}+0\ i $  \\
\hline
$ \eta^{0} T^{2/3} T^{2/3} $  & $ -2_\cdot 79\times 10^{-8} + 0 i$  \\
\hline
$ W T^{5/3} T^{5/3} $  & $ -6_\cdot 37 \times 10^{-5} + 0 i$  \\
\hline
$W' T^{5/3} T^{5/3}  $  & $ -8_\cdot 63 \times 10^{-8} + 0\ i$  \\
\hline
$ H^{\pm} T^{5/3} T^{5/3} $  & $ 8_\cdot 84 \times 10^{-9} +0\ i$  \\
\hline
\end{tabular}
%\caption{BLHM contributions to $a^{W}_{t}$.
%\label{CN2}}
\end{table}

\begin{table}[H]
\caption{Continuation of Table~\ref{CN2}.
\label{CN3}}
\centering
\begin{tabular}{|c|c|}
\hline
\hline
\multicolumn{2}{|c|}{$\sqrt{q^{2}}=500\ \text{GeV}$,\ $\text{m}_{A_0}=1 000\ \text{GeV}$, $\text{m}_{\eta_0}=100\ \text{GeV}$, $f=1 000\ \text{GeV}$, $F=4000\ \text{GeV}$}\\
\hline
\hline
$W BB $  & $ 8_\cdot 53 \times 10^{-5} +0\ i$  \\
\hline
$ W' BB $  & $ 1_\cdot 18 \times 10^{-7} +0\ i$  \\
\hline
$H^{\pm} BB $  & $ -2_\cdot 82\times 10^{-8} + 0\ i$  \\
\hline
$ \textbf{Total} $  & $ 2_\cdot 49 \times 10^{-4} - 1_\cdot 26 \times 10^{-5}\ i $  \\
\hline
\end{tabular}
%\caption{BLHM contributions to $a^{W}_{t}$.
%\label{CN2}}
\end{table}

\newpage

\end{document}